\documentclass[12pt,preprint]{aastex}

\shorttitle{The Globular Cluster System of M60. II.}
\shortauthors{Hwang et al.}

\begin{document}
\title{The Globular Cluster System of M60 (NGC 4649).
II. Kinematics of the Globular Cluster System\altaffilmark{*}}

\author{Ho Seong Hwang\altaffilmark{1,5,6}, Myung Gyoon Lee\altaffilmark{1},
Hong Soo Park\altaffilmark{1}, Sang Chul Kim\altaffilmark{2}, \\
Jang-Hyun Park\altaffilmark{2}, Young-Jong Sohn\altaffilmark{3},
Sang-Gak Lee\altaffilmark{1}, Soo-Chang Rey\altaffilmark{4}, \\
Young-Wook Lee\altaffilmark{3}, Ho-Il Kim\altaffilmark{2}}

\altaffiltext{1}{Astronomy Program, Department of Physics and Astronomy,
Seoul National University, 56-1 Sillim 9-dong, Gwanak-gu, Seoul 151-742, Korea}

\altaffiltext{2}{Korea Astronomy and Space Science Institute,
Daejeon 305-348, Korea}

\altaffiltext{3}{Center for Space Astrophysics, Yonsei University,
Seoul 120-749, Korea}

\altaffiltext{4}{Department of Astronomy and Space Sciences,
Chungnam National University, Daejeon 305-764, Korea}

\altaffiltext{5}{hshwang@kias.re.kr}

\altaffiltext{6}{Current address: Korea Institute for Advanced Study, Seoul 130-722, Korea}

\altaffiltext{7}{*Based on observations obtained at the Canada-France-Hawaii
Telescope (CFHT) which is operated by the National Research
Council of Canada, the Institut National des Science de l'Univers
of the Centre National de la Recherche Scientifique of France,
and the University of Hawaii.
This work is organized by the Korea Astronomy and Space Science Institute.}

\begin{abstract}
We present a kinematic analysis of the globular cluster (GC) system in
the giant elliptical galaxy (gE) M60 in the Virgo cluster.
Using the photometric and spectroscopic database of 121 GCs (83 blue GCs and 38 red GCs),
we have investigated the kinematics of the GC system.
We have found that the M60 GC system shows a
significant overall rotation. The rotation amplitude of the blue GCs is
slightly smaller than or similar to that of the red GCs,
and their angles of rotation axes are similar.
The velocity dispersions about the mean velocity and about
the best fit rotation curve for the red GCs are marginally larger
than those for the blue GCs.
Comparison of observed stellar and GC velocity dispersion profiles with those
calculated from the stellar mass profile
shows that the mass-to-light ratio should be increased as the galactocentric distance increases,
indicating the existence of an extended dark matter halo.
The entire sample of GCs in M60 is found to have a
tangentially biased velocity ellipsoid unlike the GC systems in other gEs.
Two subsamples appear to have different velocity ellipsoids.
The blue GC system has a modest tangentially biased velocity ellipsoid, while
the red GC system has a modest radially biased or an isotropic velocity ellipsoid.
From the comparison of the kinematic properties of the M60 GC system to those of other gEs
(M87, M49, NGC 1399, NGC 5128, and NGC 4636),
it is found that the velocity dispersion
of the blue GC system is similar to or larger than that of the red GC system except for M60, and
the rotation of the GC system is not negligible. The entire sample
of each GC system shows an isotropic velocity ellipsoid except for M60, while the
subsamples show diverse velocity ellipsoids. We discuss 
the implication of these results for the formation models of the GC system in gEs.
\end{abstract}



\section{Introduction}
Globular Clusters (GCs) have been long recognized as an important
tracer to understand the formation and evolution of galaxies. To
solve the mystery of galaxy formation and evolution, several
photometric properties of the GCs such as color distribution,
spatial structure, and luminosity function have been used
\citep{lee03,west04,brodie06}. However, it is still difficult to
test the predictions of the formation models of galaxies and their
GC systems using photometric data alone.

Recently, with an aid of large telescopes with apertures larger than 4 meter,
  a large sample ($N>150$) of GC spectra for one galaxy has been obtained, which can be used for
  statistically meaningful kinematic studies (e.g., \citealt{cote01,coteetal03,richtler04,peng04}).
Due to their wide spatial distribution, brightness, and compact size,
  GCs are useful test particles to trace the gravitational potential of their host galaxies,
  and especially of the dark matter halo beyond several effective radii of the galaxy.
Therefore, a kinematic study of the GC system
  enables us to estimate the global mass distribution of their host
  galaxy, or to constrain the orbital properties of GCs
  using an independently determined mass profile (e.g., from X-ray emission) of the galaxy.
Moreover, the kinematic difference between GC subpopulations (blue and red GCs)
  can be used as an observational constraint on the
  galaxy formation model.

To date there are six giant elliptical galaxies (gEs) for
  which the kinematics of the GC system have been studied to our knowledge:
  M87 \citep{cohen97,kissler98,cote01}, M49 \citep{zepf00,coteetal03},
  NGC 1399 \citep{kissleretal98,minniti98,kissler99,richtler04},
  NGC 5128 \citep{peng04,wood07}, NGC 4636 \citep{schuberth06}, and M60 \citep{bri06}.
M87, cD galaxy of the Virgo cluster, was studied recently by \citet{cote01} using $\sim280$ velocity data of GCs.
  The velocity dispersions of the blue and red GCs were found to be $\sim$410 km s$^{-1}$ and
  $\sim$390 km s$^{-1}$, respectively. Both the blue and red GCs appear to rotate around the photometric minor axis
  with a similar rotation amplitude of $\sim160$ km s$^{-1}$ when averaged over the whole system, while the
  blue GCs appear to rotate around the photometric major axis inside a radius of $\sim 16$ kpc. The
  entire GC system has an isotropic velocity ellipsoid, while the blue and red GC systems show tangentially and
  radially biased one, respectively.
\citet{coteetal03} studied the GC kinematics of M49, the brightest member of the Virgo cluster, using $\sim260$ velocity data of GCs.
  They found a larger velocity dispersion of the blue GCs ($\sim$350 km s$^{-1}$) than that of the red GCs ($\sim$270 km s$^{-1}$).
  The blue GCs rotate roughly around the photometric axis of M49 with a rotation amplitude of $\sim100-150$ km s$^{-1}$,
  while the red GCs show some evidences for weak rotation ($\sim50$ km s$^{-1}$) around the same axis but a different direction.
  The entire, blue, and red GC systems appear to have an isotropic velocity ellipsoid.
NGC 1399, cD galaxy of the Fornax cluster, was studied by \citet{richtler04} with the largest number ($\sim470$) of GC velocity data.
  The velocity dispersions of the blue and red GCs were estimated to be $\sim$291 km s$^{-1}$ and $\sim$255 km s$^{-1}$, respectively.
  No significant rotations were found for both the blue and red GCs, while there is a weak signature of rotation
  for blue GCs beyond 6$\arcmin$. The velocity anisotropies for both the blue and red GCs are consistent with isotropic orbits.
Using $\sim220$ GC data of NGC 5128, \citet{peng04} found that the red GCs exhibit a significant rotation,
  while the blue GCs do not show a clear hint of rotation. Later, \citet{wood07}, using 340 GC data for NGC 5128,
  showed that the rotation amplitude and the velocity dispersion for the subsamples are quite similar.
For NGC 4636, \citet{schuberth06} found that the velocity dispersions of the blue and red GCs are not different,
  but the rotation of the red GCs is stronger than that of the blue GCs.
In summary, all these GC systems show diverse GC kinematics in velocity dispersion, rotation, and
  their radial variation (see Section \ref{comp_ge}),
  making it difficult to draw any strong conclusion on the uniform formation
  history of the GC system in gEs.

M60 (NGC 4649) is a giant elliptical galaxy in the Virgo cluster, slightly less luminous than M87 and M49.
M60 has a nearby companion Sc galaxy, NGC 4647, located at $2.\arcmin5$ from the center of M60.
While there were numerous photometric studies of the GC system in M60 based on the
  ground-based images \citep{couture91,harris91,ashman98,forbes04,lee07} and
  Hubble Space Telescope images \citep{neilsen99,kundu01,larsen01,peng06,strader05,mie06},
  there were few spectroscopic studies of the M60 GC system \citep{pierce06}.
\citet{pierce06} published for the first time the spectroscopic observational results
  for 38 GCs (16 blue and 22 red GCs) in M60 using Gemini/GMOS.
They found no obvious signs of a recent starburst,
  interaction or merger by estimating the ages and metallicities from
  the spectra of M60 GCs.
Later, \citet{bri06} investigated the kinematics of M60 GC system using the velocity data of \citet{pierce06}.
They reported that the velocity dispersion of the blue GCs is smaller than
  that of the red GCs unlike the cases of other gEs.
They found no hint of rotation in the GC system of M60.
Furthermore, orbital distribution of the GC system is found to be close to isotropic
  inside the radius of 100$\arcsec$, but becomes tangentially biased beyond this radius.
However, due to their small number ($N\sim38$) of GC velocity data with
  limited radial and azimuthal coverage,
  it was difficult to derive the kinematic properties up to several effective radii,
  and to distinguish the kinematic difference between the blue and red GCs.

We have carried out a photometric study of the GC system in M60 using deep wide-field $CT_1$ images
  obtained at the KPNO 4m telescope \citep{lee07}, and a spectroscopic study
  using the spectra obtained at the Canada-France-Hawaii Telescope (CFHT) \citep[Paper I]{lee06}.
In this paper, we present the results of kinematic study of the GC system in
M60 using the velocity data of 121 GCs selected in Paper I.
Section 2 gives a brief description of the data used in this analysis, and the kinematic properties
are derived in Section 3.
From the comparison of kinematic properties of the M60 GC system to those of other gEs,
we discuss our results regarding the GC formation models in Section 4.
A summary of this study is given in the final section.
We adopt a distance to M60, 17.3 Mpc, given by \citet{mei07} based on a surface brightness fluctuation method,
  and the corresponding scale for one arcmin is 5.032 kpc. 

\section{Data}

We used the spectroscopic data of GCs given in Paper I,
which describes the details of the spectroscopic observation,
data reduction, and the data set. Here we only give a brief summary of the data set of M60 GCs.

We selected GC candidates in deep, wide-field Washington
C and $T_1$ images ($16\arcmin\times16\arcmin$) obtained at KPNO 4m telescope \citep{lee07}
  and F555W (V) and F814W (I) images in the HST/WFPC2 archive data
  \citep{neilsen99,kundu01,larsen01}.
Spectroscopic observations were made using the Multi Object
Spectrograph (MOS) at the 3.6 m CFHT in February 2002 and in May 2003 for 165 GC candidates with $19<T_1<22 \ \mathrm {mag}$ and
$1.0\le(C-T_1)<2.4$. We determined the radial velocities of GC candidates by cross-correlating
the candidate spectrum with that of three Galactic GCs.
Among 165 GC candidates, we could extract the spectra
and determine the radial velocities for 111 objects, except for 54 objects due to the poor quality of their spectra.
We increased the number of GCs by combining our data of radial velocities with those of \citet{pierce06}.
The radial velocities of GCs in \citet{pierce06} were transformed
into our velocity system using equation (1) in Paper I,
and the transformed velocities were used for further analysis.
Of the entire spectroscopic sample of GC candidates (110 from Paper I and 38 from \citealt{pierce06}),
121 genuine GCs were selected in Paper I using a criteria of radial velocities ($500 \le v_p \le 1600~{\rm
km~s}^{-1} $) and ($C-T_1$) colors ($1.0 \le C-T_1 < 2.4~{\rm mag}$).
There are 83 blue GCs with $1.0\le(C-T_1)<1.7$ and 38 red GCs with $1.7\le(C-T_1)<2.4$ in the total sample.
Foreground reddening toward M60 is very small,
$E(B-V)=0.026$ \citep{sch98}, corresponding to $E(C-T_1)=1.966 E(B-V)=0.051$, $A(T_1)=0.071$, and $A(V)=0.088$.

\begin{figure}
\plotone{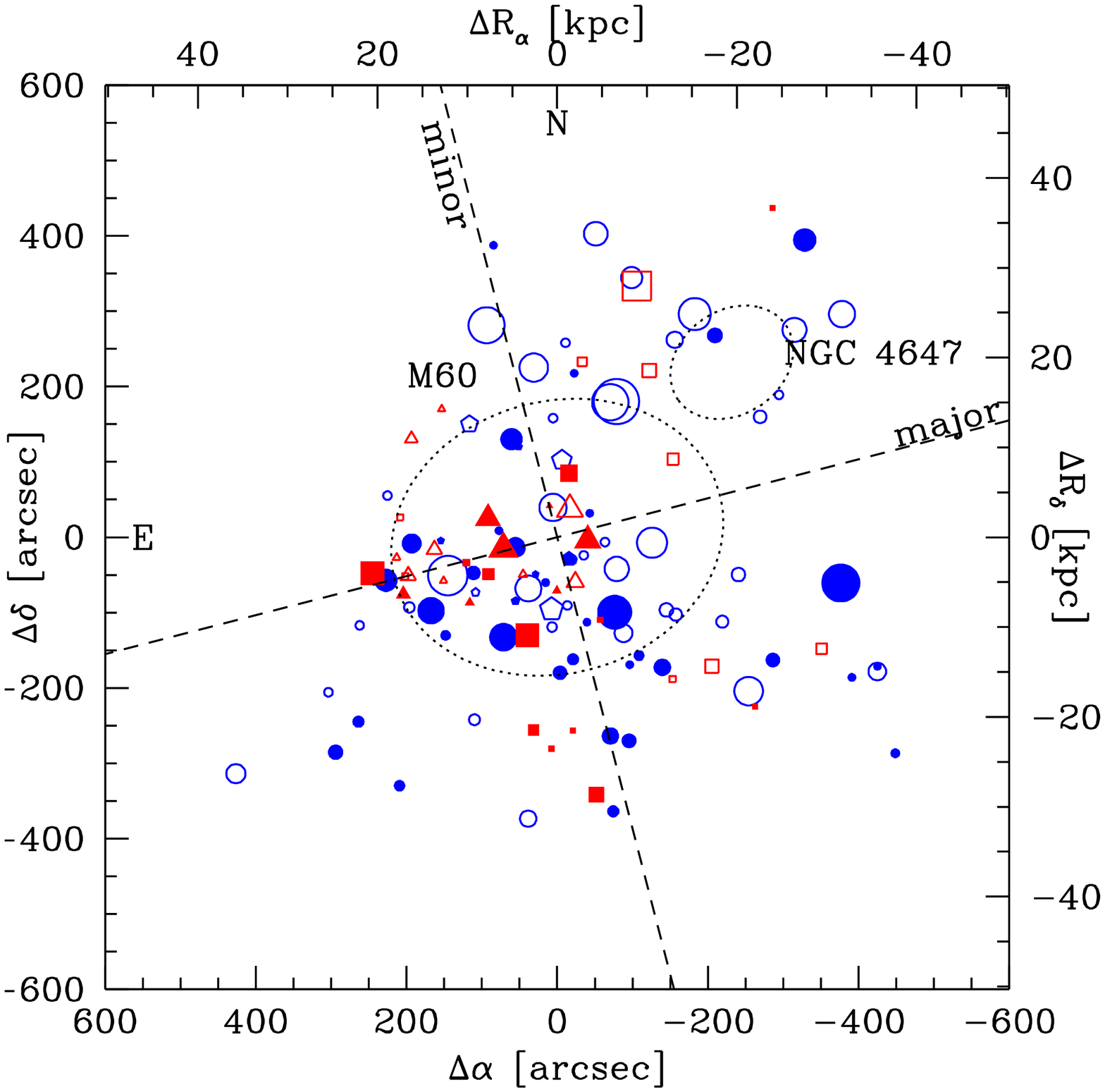} \caption{Spatial distribution of M60 GCs with measured velocities. The blue GCs measured in Paper I
  are represented by circles and those from \citet{pierce06}
  are represented by pentagons. The squares and triangles indicate
  the red GCs measured in Paper I and \citet{pierce06}, respectively.
The GCs with larger velocities than
  the systemic velocity (v$_{\rm gal}$ = 1056 km s$^{-1}$) of M60 are plotted
  by open symbols, while those with smaller velocities than
  the systemic velocity of M60 by filled symbols. The symbol size is
  proportional to the velocity deviation.
M60 and its companion galaxy, NGC 4647, are shown by large dotted $D_{25}$ ellipses.
The photometric major and minor
  axes of M60 are represented by the dashed lines. \label{fig-vmap}}
\end{figure}

In Figure \ref{fig-vmap}, we show the spatial distribution of 121 GCs
  in M60 with measured velocities.
Two large dotted ellipses represent isophotes at the 25.0 B-mag arcsec$^{-2}$ of
  M60 (larger one) and NGC 4647 (smaller one) \citep{devaucouleurs91}.
It is worth noting
  that the majority of high velocity GCs (open symbols) with larger velocities
  than the systemic velocity of M60 ($v_{\rm gal}\footnotemark\footnotetext{$v_{\rm gal}=1117\pm6$ km s$^{-1}$
  in \citet{gonzalez93}.}=1056\pm64$ km s$^{-1}$, Paper I), appear to be located to the
  north-west of M60.
On the other hand, the majority of low velocity GCs (filled symbols)
  with smaller velocities than the systemic velocity of M60
  appear to be located to the south-east of M60.
It might be related to the overall rotation of the M60 GC system around the minor axis.
However, a careful analysis is needed for further dynamical investigation due to the
  non-uniform spatial coverage of observed genuine GCs around M60.

\section{Results}

Using the master catalog of 121 GCs in M60 (Paper I),
we have investigated the kinematic properties of the M60 GC system:
the rotation amplitude, the position angle of the rotation
axis, the mean line-of-sight velocity, the projected velocity
dispersion, and the velocity ellipsoid.

\subsection{Rotation of the Globular Cluster System}

A detailed description of the relation between
the intrinsic rotational velocity field and the projected one is given in
\citet{cote01}. In summary, if we assume that the GC system is
spherically symmetric with an intrinsic angular velocity field
stratified on spheres, and that the GC rotation axis lies in the
plane of the sky, then we can find that radial velocities of GCs
depend sinusoidally on the azimuthal angles.
For the M60 GC system, the assumption of spherical symmetry of the GC system is reasonable
  due to the modest projected ellipticity (effective ellipticity, $e_{\rm eff}=0.21$, \citealt{lee07}).
Therefore, we fit the observed line-of-sight velocities ($v_p$) of the GCs with the function,

\begin{equation}
v_p (\Theta) = v_{\rm sys} + (\Omega R) \sin(\Theta - \Theta_0)\ ,
\label{eq2}
\end{equation}

where $\Theta$ is the projected position angle of GCs relative to
  the galaxy center measured from north to east, $\Theta_0$ is the
  projected position angle of the rotation axis of the GC system, 
  $R$ is the projected galactocentric distance,
  ($\Omega R$) is the rotation amplitude, and $v_{\rm sys}$ is the systemic velocity of the GC system.

\begin{figure}
\plotone{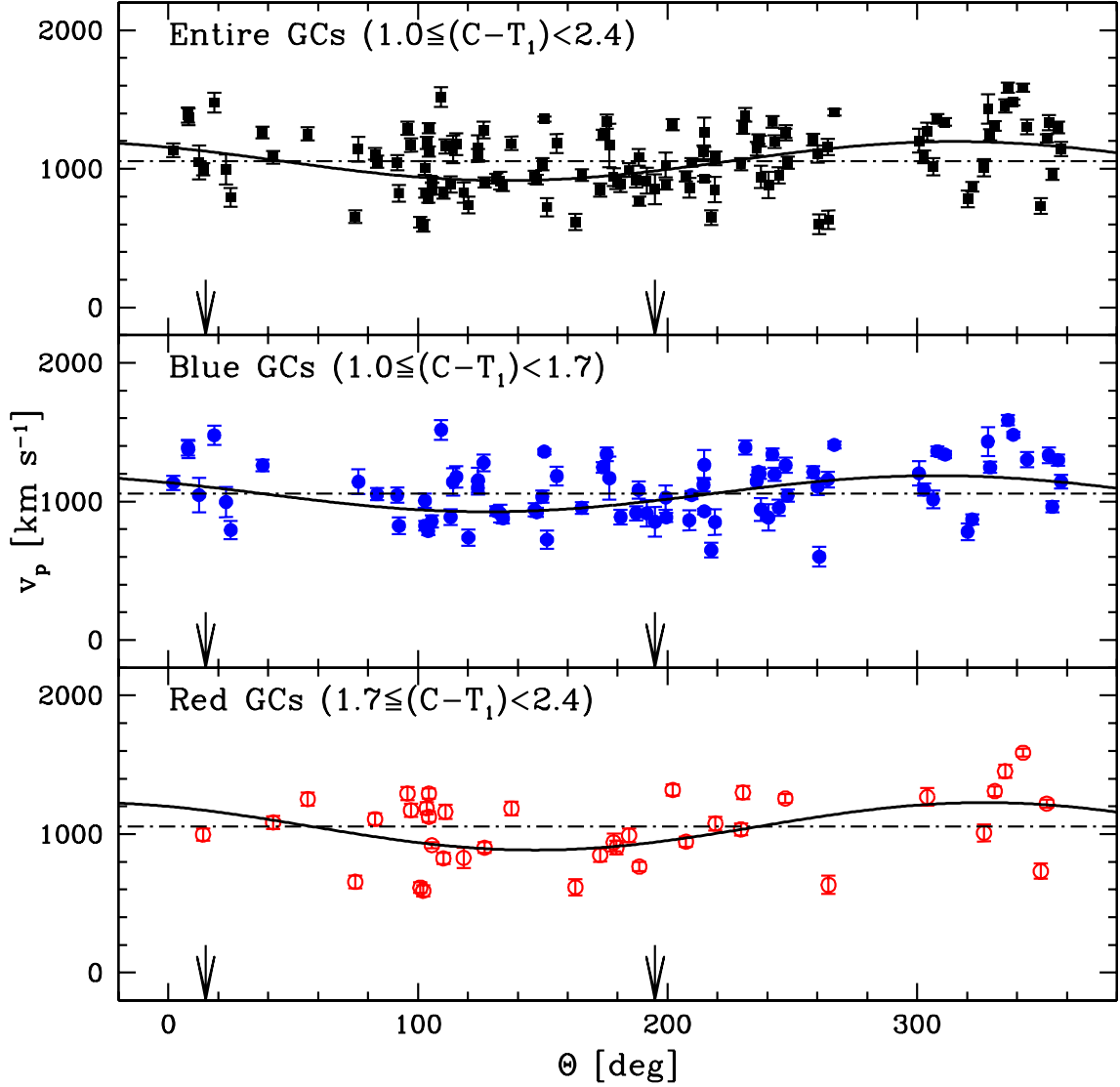} \caption{Radial velocities versus position angles
for entire 121 GCs (top panel), 83 blue GCs (middle panel), and 38
red GCs (bottom panel). The solid curve represents the best fit
rotation curve from Table~\ref{tab-m60kin}, and the dot-dashed
horizontal line indicates the systemic velocity of M60. The
photometric minor axis of M60 is represented by the vertical
arrows ($\Theta = 15^{\circ}$ and $195^{\circ}$).
\label{fig-rotsub}}
\end{figure}

In Figure~\ref{fig-rotsub}, we plot the radial velocities of GCs with
  measured uncertainties as a function of position angle
  for entire 121 GCs (top panel), 83 blue GCs (middle panel), and 38 red GCs (bottom panel).
The best fit rotation curve of equation (\ref{eq2}) for each sample is overlaid.
The fitting was done using an error-weighted, nonlinear fit of equation (\ref{eq2})
  with $v_{\rm sys}$ as a fixed value of M60 recession velocity
  ($v_{\rm gal}=1056\pm64$ km s$^{-1}$) rather than a free parameter for a better fitting.
Using the biweight location of \citet{beers90}, the systemic velocity of the M60 GC
  system is estimated to be $v_{\rm sys}=1073^{+22}_{-22}$ km s$^{-1}$ (see Section \ref{dispersion}),
  which agrees with the M60 recession velocity (1056$\pm64$ km s$^{-1}$) and
  the biweight mean velocity (1066$\pm$45 km s$^{-1}$) for 38 GCs given by \citet{bri06} within the uncertainty.
We derived the rotation amplitudes, $\Omega R$,
  $141^{+50}_{-38}$ km s$^{-1}$ for the entire GCs,
  $130^{+62}_{-51}$ km s$^{-1}$ for the blue GCs, and
  $171^{+58}_{-46}$ km s$^{-1}$ for the red GCs.
Thus, the rotation amplitude of the blue GCs is slightly smaller than that of the red GCs,
  or is consistent with that of the red GCs within the uncertainty.
Our results based on 121 GCs are in contrast to the fact that
  \citet{bri06} found no rotation using only 38 GCs.
To investigate the cause for difference of the rotation amplitudes between this study
  and \citet{bri06}, we estimate the rotation amplitude
  using all 121 GCs by dividing GCs into 54 GCs within the region of \citet{bri06},
  and 67 GCs outside the region of \citet{bri06}.
In the result, the rotation amplitude for the former sample
  is estimated to be $74^{+90}_{-27}$ km s$^{-1}$,
  which is much smaller than that for the latter sample ($\Omega R=221^{+42}_{-41}$ km s$^{-1}$).
Therefore, it is concluded that no rotation found in \citet{bri06} is because of a small spatial coverage
  in their study.

In addition, the orientation of the rotation axis ($\Theta_0$) is estimated to be
  ${225^\circ}^{+12}_{-14}$ for the entire GCs,
  ${218^\circ}^{+16}_{-23}$ for the blue GCs, and
  ${237^\circ}^{+18}_{-19}$ for the red GCs.
The orientations of rotation axes for all subsamples appear to be similar,
  and they are closer to the photometric minor axis
  ($\Theta_{phot} = 15^\circ$, or $195^\circ$, \citealt{lee07}) than the photometric
major axis. Interestingly, the position angle of NGC 4647 relative to M60 is 314$^\circ$ from north to east.
This means that the line connecting M60 and NGC 4647 is nearly
  perpendicular to the rotation axes of the GC system.

\begin{figure}
\plotone{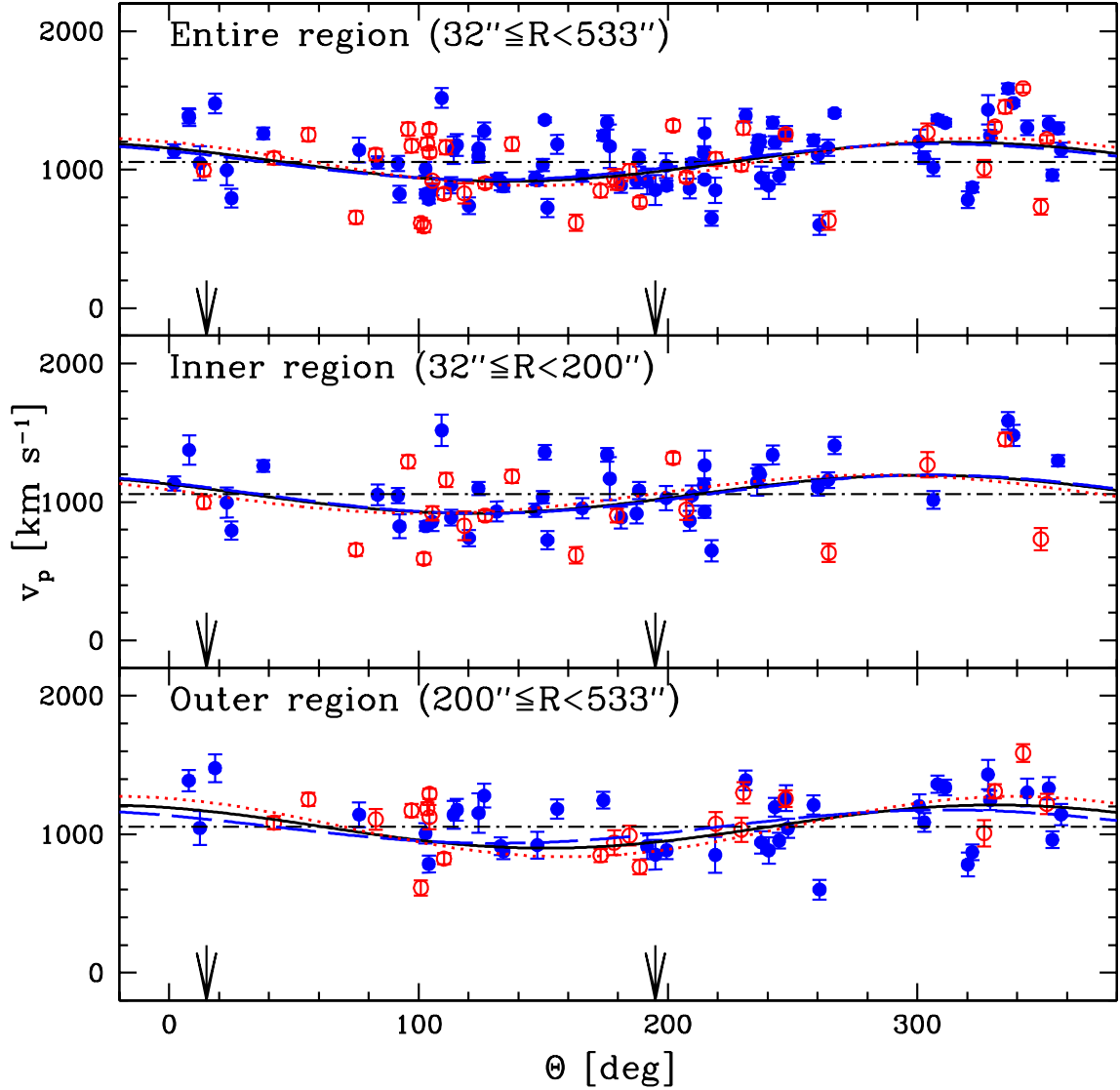} \caption{Radial velocities versus position angles
for the GCs in the range of $32\arcsec \leq R < 533\arcsec$ (top
panel), $32\arcsec \leq R < 200\arcsec$ (middle panel), and
$200\arcsec \leq R < 533\arcsec$ (bottom panel). Filled
circles indicate the blue GCs, while open circles the red GCs.
The best fit rotation curves for the entire (solid curves),
blue (dashed curves) and red (dotted curves) GCs
within each region are overlaid. The dot-dashed horizontal line
indicates the systemic velocity of M60, and the vertical arrows
mark the position angle of photometric minor axis of M60.
\label{fig-rotreg}}
\end{figure}

In Figure~\ref{fig-rotreg}, we present the rotation of the GC
  system for the samples of different radial bins in order to investigate
  the radial variation of rotational properties. The top panel shows
the GCs in the range of $32\arcsec \leq R < 533\arcsec$, while the
lower two panels show the GCs for the inner region ($32\arcsec
\leq R < 200\arcsec$, middle panel) and the outer region
($200\arcsec \leq R < 533\arcsec$, bottom panel).
The same fitting procedure used for the results in Figure~\ref{fig-rotsub} was applied
  for the entire, blue, and red GCs in each radial bin.
The best fit rotation curves for each
  sample are overlaid as solid (entire GCs), dashed (blue GCs), and dotted (red GCs) lines.

It appears that the orientation of rotation axes for all GC
  subsamples changes slightly from the inner region to the outer region
  with a large uncertainty.
The change of the rotation axis from the inner region to
the outer region for the blue GCs is seen in the cases of M87
\citep{cote01} and of M49 \citep{coteetal03}. The orientation of the
rotation axis for the blue GCs in M87 drastically changes from the
major axis in the inner region to the minor axis in the outer region.
However, the blue GCs in M49 show the
  opposite change of rotation axis from the minor axis in the inner
  region to the major axis in the outer region.
Due to the limited number of GCs in this study,
  it is difficult to make a solid conclusion on the radial change of the rotation
  axis for the M60 GC system.
A larger sample of GCs is needed for further study on the
  change of the rotation axis as a function of the radius.

\subsection{Velocity Dispersion of the Globular Cluster System}\label{dispersion}

We summarize the kinematics of the M60 GC system derived in this study in Table~\ref{tab-m60kin}. 
Several kinematic parameters for the entire, blue, and red GCs are presented 
 for the entire region ($32\arcsec \leq R < 533\arcsec$),
the inner region ($32\arcsec \leq R < 200\arcsec$), and the outer region ($200\arcsec \leq R < 533\arcsec$).
The column (1) defines the range of the projected radial distance from the center of M60
  for each region in arcsec, and the column (2) gives the median value of
  the radial distance in arcsec.
The number of GCs in each region is shown in the column (3). The column (4) and (5) represent the mean
line-of-sight velocity (the biweight location of \citealt{beers90})
and the velocity dispersion about this mean velocity (the biweight scale of \citealt{beers90}), respectively. The position
angle of the rotation axis and the rotation amplitude estimated using
equation (\ref{eq2}) in each region are given in the column (6) and (7), respectively.
The column (8) gives the velocity dispersion
  about the best fit rotation curve. The column (9) gives the absolute value
  of the ratio of the rotation amplitude to the velocity dispersion about the
best fit rotation curve. The uncertainties of these values
represent $68\%$ $(1\sigma)$ confidence intervals that are
determined from the numerical bootstrap procedure following the method of \citet{cote01}.
The estimated velocity dispersion for entire 121 GCs ($234^{+13}_{-14}$ km s$^{-1}$) agrees with the value ($256\pm29$ km s$^{-1}$) of \citet{bri06}
based on 38 GCs within the uncertainty.
In addition, it is found that the velocity dispersion about the mean velocity of the GC system
for the red GCs ($\sigma_p = 258^{+21}_{-31}$ km s$^{-1}$) is marginally larger than
that for the blue GCs ($\sigma_p = 223^{+13}_{-16}$ km s$^{-1}$),
which confirms the result of \citet{bri06} with improved statistics.
Interestingly, the velocity dispersion about the best fit rotation curve
  for the red GCs ($\sigma_{p,r}=240^{+20}_{-34}$ km s$^{-1}$) is also marginally larger than
  that for the blue GCs ($\sigma_{p,r}=207^{+15}_{-19}$ km s$^{-1}$).

\begin{figure}
\plotone{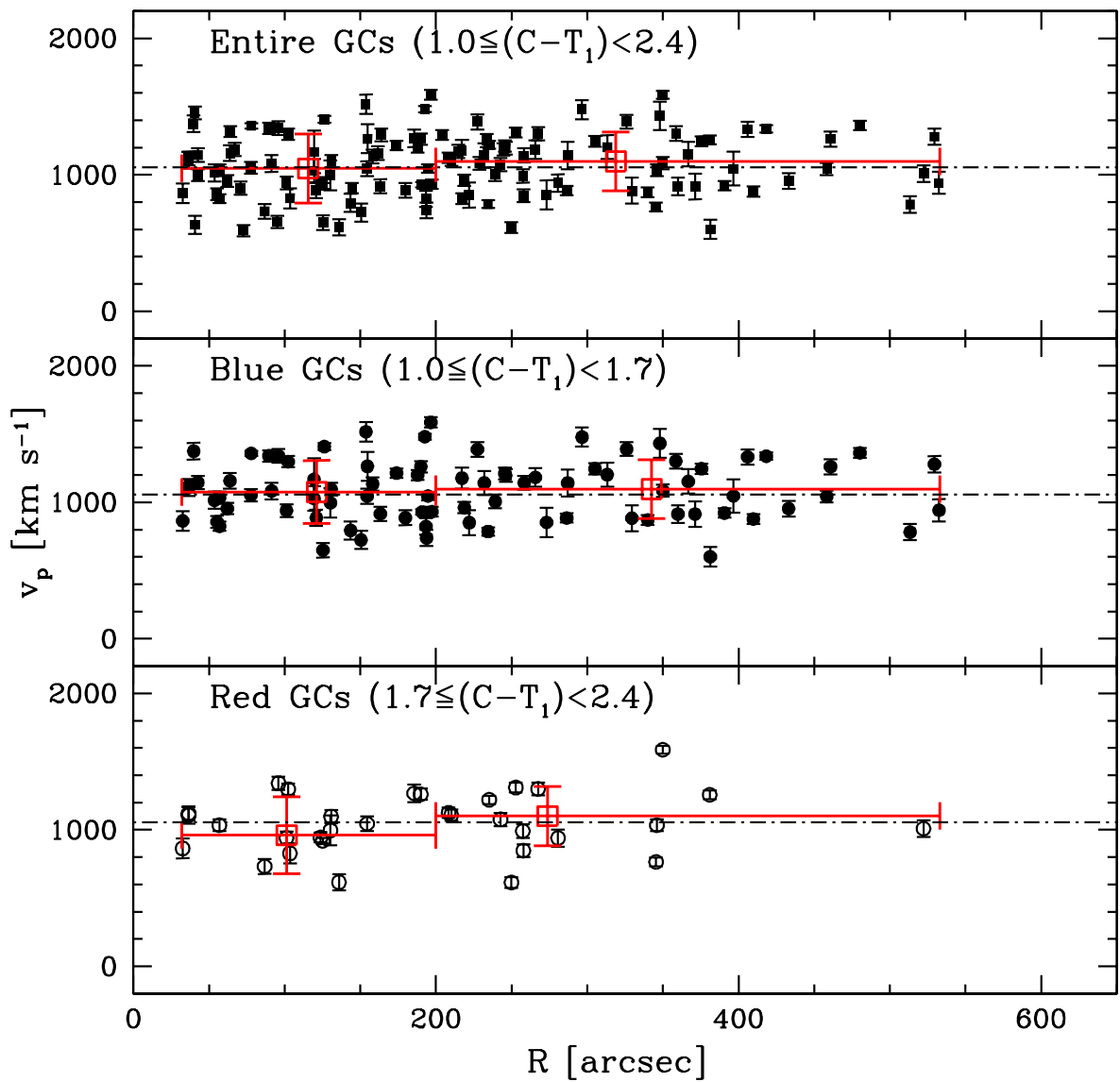} \caption{Radial velocities versus projected
  galactocentric distances for the entire (top panel), blue (middle panel),
  and red (bottom panel) GCs. Large open squares
indicate the mean radial velocities of GCs in the radial bins that
are represented by long horizontal errorbars. Their vertical
errorbars denote the velocity dispersions of GCs in the radial
bins. The dot-dashed horizontal line indicates the systemic
velocity of M60. \label{fig-vel}}
\end{figure}

In Figure~\ref{fig-vel}, we plot the radial velocities of GCs with
measured uncertainties against projected galactocentric distances. The
mean radial velocities in two radial bins are overlaid by squares with long horizontal errorbar.
The velocity dispersion about the mean velocity in each bin is also represented
  by a vertical errorbar. The mean velocities of all samples agree
well with the systemic velocity of M60. Both of the velocity dispersions
about the mean velocity and about the best fit rotation curve in
the inner region are marginally larger than or comparable to those in the outer
region for all three samples (see Table~\ref{tab-m60kin}).

\begin{figure}
\plotone{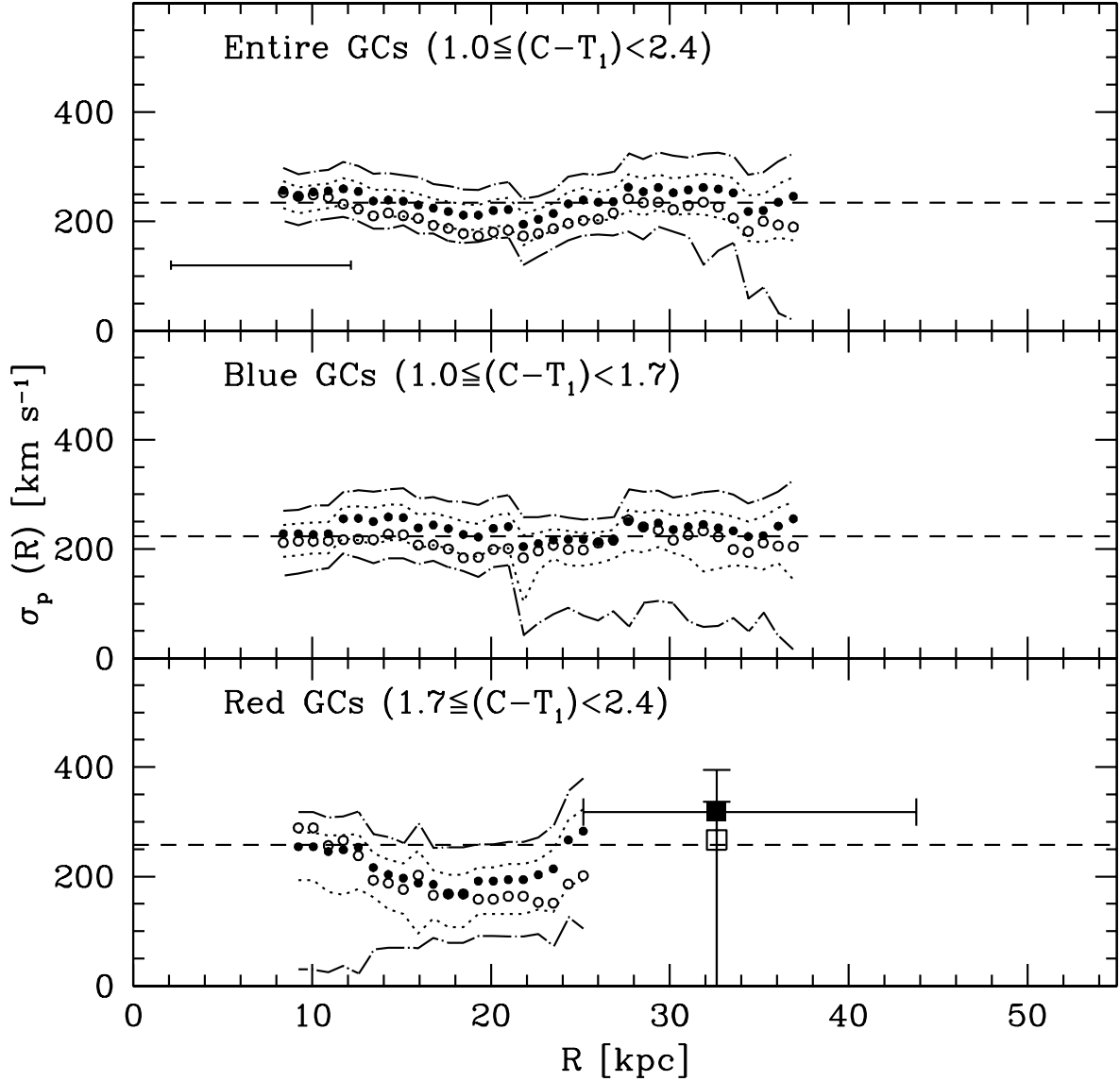} \caption{Radial profiles of velocity dispersion
for the entire (top panel), blue (middle panel), and red
(bottom panel) GCs. Filled circles indicate the velocity
dispersion about the mean GC velocity
($\sigma_p$) at each point, while open circles the velocity
dispersion about the best fit rotation curve ($\sigma_{p,r}$) at
the same point. The dispersion is calculated using the GCs within
the moving radial bin (width of $2\arcmin\simeq 9.77$ kpc) that is
represented by a horizontal errorbar in top panel. The dotted and
dot-dashed lines denote 68\% and 95\% confidence intervals on the
calculation of velocity dispersion, respectively. The dashed
horizontal line indicates the global value of velocity dispersion
of GCs in each panel. In the bottom panel, large filled and open squares, respectively,
represent the velocity dispersion about the mean GC velocity ($\sigma_p$)
and that about the best fit rotation curve ($\sigma_{p,r}$) for the red GCs beyond $R\simeq25.1$ kpc. \label{fig-disp}}
\end{figure}

To investigate the radial variation of velocity dispersion in detail,
  we present a smoothed radial profile of velocity
  dispersions about the mean radial velocity (filled symbols) and about the
  best fit rotation curve (open symbols) in Figure~\ref{fig-disp}.
We calculate the velocity dispersion of the GCs lying within a bin with fixed radial width,
  $\Delta R = 120\arcsec \simeq 10.06$ kpc as increasing the
  bin center by a fixed step width, $\delta R = 10\arcsec\simeq 0.84$ kpc.
We define the radial width and the step width so
  that the number of GCs per bin exceeds 10, and the calculation stops
  when the number of GCs in a bin is less than 10.
The velocity dispersions about the mean radial velocity of the entire and blue GCs
  are nearly constant for the range of radius.
However, velocity dispersions of the red GCs are decreasing in the inner region
($R\lesssim 18$ kpc) and are increasing in the outer region ($R\gtrsim
18$ kpc) although the variation is not large. The velocity
dispersions about the best fit rotation curves of all three samples
are not different from those about the mean radial velocities.

\subsection{Velocity Anisotropy of the Globular Cluster System}

If we assume the spherical symmetry of the M60 GC system, we
can apply the Jeans equation in the absence of rotation to the
dynamical analysis of the GC system. The spherical Jeans equation is

\begin{equation}
{d\over{dr}}\, n_{\rm cl}(r) \sigma_r^2(r) + {{2\,\beta_{\rm
cl}(r)}\over{r}}\, n_{\rm cl}(r) \sigma_r^2(r) = - n_{\rm
cl}(r)\,{{G M_{\rm tot}(r)}\over{r^2}}\ , \label{eq4}
\end{equation}

where $r$ is a three dimensional radial distance from the galactic center,
  $n_{\rm cl}(r)$ is a three dimensional density profile of the GC system,
  $\sigma_r(r)$ is a radial component of velocity dispersion,
  $\beta_{\rm cl}(r)\equiv1-\sigma_\theta^2(r)/\sigma_r^2(r)$ is a velocity anisotropy,
  $G$ is the gravitational constant,
  and $M_{\rm tot}(r)$ is a total gravitating mass contained
  within a sphere of radius $r$ (e.g., \citealt{bt87}).
$\sigma_\theta(r)$ is a tangential component of velocity dispersion that is
  equal to an azimuthal component of the velocity dispersion, $\sigma_\phi(r)$,
  in a spherical case.

Several studies on the dynamics of the GC system have focused on
  determining the gravitational mass, $M_{\rm tot}(r)$, using the Jeans
  equation by assuming a simple isotropic orbit with $\beta_{\rm
  cl}(r)=0$ (e.g., \citealt{cohen97,minniti98,zepf00}).
However, with an aid of an independent determination of the mass profile of an
  elliptical galaxy using X-ray data (e.g., \citealt{brighenti97,hum06} for M60),
  the velocity anisotropy itself can be investigated (e.g., \citealt{romanowsky01,cote01,coteetal03}).
Following the analysis of the M87 GC system by \citet{cote01} and the M49 GC
system by \citet{coteetal03}, we derive first the three dimensional
density profile of the GC system, $n_{\rm cl}(r)$ and the total mass
profile, $M_{\rm tot}(r)$.
Comparing the velocity dispersion profile (VDP) calculated from the Jeans equation
  using those inputs with the measured VDP $\sigma_p(R)$, we determine
  the velocity anisotropy of the M60 GC system. 

\subsubsection{Density Profiles for the GC system}
We used the surface number density profiles of M60 GCs in \citet{lee07}.
They derived the surface density profile
  of M60 GCs by combining the HST/WFPC2 archive data
  for the inner region at $R<1.5\arcmin$,
  and the KPNO data for the outer region at $R>1.5\arcmin$.
They determined the background levels from the mean surface number density
  of the point sources at $9\arcmin-10\arcmin$
  with the same range of magnitude and color as the GCs in the KPNO images:
  $1.988\pm 0.363$ per square arcmin for the entire GCs,
  $1.790\pm 0.344$ per square arcmin for the blue GCs,
  and  $0.199\pm 0.115$ per square arcmin for the red GCs.
Then they subtracted these background values from the original number counts
  to produce the radial profiles of the net
  surface number density of GCs.
Since they selected GCs that are brighter than $T_1 \approx 23.0$
  mag\footnotemark\footnotetext{This limiting magnitude corresponds
  to $V \approx 23.31$ mag using the transformation relation between
  $VI$ system and $CT_1$ system of \citet{lee07}.},
  it is needed to correct the surface number density
  profile in order to account for the uncounted GCs due to the
  limiting magnitude.
To calculate the correction factor, the equation (11) in \citet{mclaughlin99} with $V_{lim,1}=\infty$ was
  used on the assumption that the GC luminosity function of M60 has
  a Gaussian shape with a peak at $V \approx 23.8$ mag and a
  dispersion $\sigma = 1.65 $ mag \citep{kundu01}.
It is found that the surface number density of the bright M60 GCs in \citet{lee07} should be multiplied by 2.61
  to derive the total surface number density.

\begin{figure}
\plotone{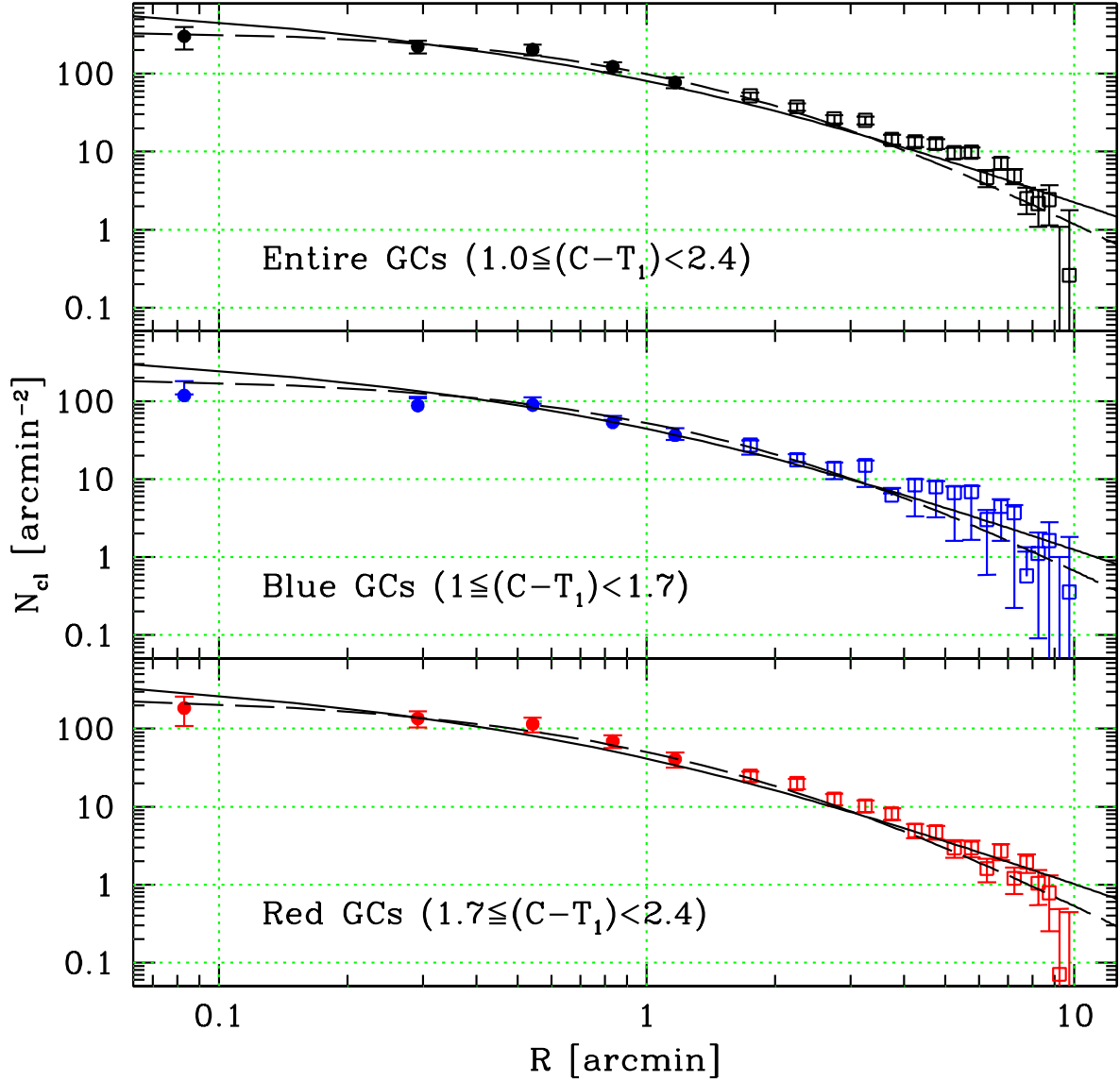} \caption{Projected number density profiles for
the entire (top panel), blue (middle panel), and red (bottom panel) GC candidates.
Filled circles represent the GC candidates from HST/WFPC2 archive,
while open squares the GC candidates from the KPNO $CT_1$ images \citep{lee07}. The solid line and the
dashed line in each panel indicate the projected best fits using
the NFW density profile and the Dehnen density profile,
respectively, for each sample. \label{fig-numden}}
\end{figure}

We display the total surface number density profiles, $N_{\rm cl}(R)$, for
  the entire, blue and red GCs in Figure \ref{fig-numden}.
We fit the surface number density profile with the projection of \citet[NFW]{nfw97}
  density profile, $n_{\rm cl} (r)=n_0(r/b)^{-1}(1+r/b)^{-2}$ and with
  the projection of one of the galaxy models developed by \citet{dehnen93},
  $n_{\rm cl} (r)=n_0(r/a)^{-\gamma}(1+r/a)^{\gamma-4}$.
The surface number density profile, $N_{\rm cl}(R)$, is
  related to the three dimensional density profile $n_{\rm cl}(r)$ as follows:
\begin{equation}
N_{\rm cl}(R) = 2\int_{R}^{\infty} n_{\rm cl}(r)
{{r\,dr}\over{\sqrt{r^2-R^2}}}\ . \label{eq7}
\end{equation}

The solid and long dashed lines represent the projected best fit curves
of the NFW profile and of the Dehnen profile, respectively.
The fitting results for the entire (E) GCs, blue (B) GCs, and red (R) GCs
are summarized as follows:
\begin{equation}
\begin{array}{rcl}
n_{\rm cl}^{\rm E}(r) & = &
  0.61\,{\rm kpc}^{-3}(r/5.96\,{\rm kpc})^{-1}(1+r/5.96\,{\rm kpc})^{-2} \\
n_{\rm cl}^{\rm B}(r) & = &
  0.32\,{\rm kpc}^{-3}(r/6.07\,{\rm kpc})^{-1}(1+r/6.07\,{\rm kpc})^{-2} \\
n_{\rm cl}^{\rm R}(r) & = &
  0.46\,{\rm kpc}^{-3}(r/4.99\,{\rm kpc})^{-1}(1+r/4.99\,{\rm kpc})^{-2},  \\
\end{array}
\label{eq-nfw}
\end{equation}
for the NFW profile, and
\begin{equation}
\begin{array}{rcl}
n_{\rm cl}^{\rm E}(r) & = &
  1.42\,{\rm kpc}^{-3}(r/7.97\,{\rm kpc})^{-0.29}(1+r/7.97\,{\rm kpc})^{-3.71} \\
n_{\rm cl}^{\rm B}(r) & = &
  0.57\,{\rm kpc}^{-3}(r/8.75\,{\rm kpc})^{-0.40}(1+r/8.75\,{\rm kpc})^{-3.60} \\
n_{\rm cl}^{\rm R}(r) & = &
  0.53\,{\rm kpc}^{-3}(r/8.36\,{\rm kpc})^{-0.56}(1+r/8.36\,{\rm kpc})^{-3.44},  \\
\end{array}
\label{eq-dh}
\end{equation}
for the Dehnen profile.

It is found that the scale length $b$ of the red GCs in the NFW profile is smaller
  than that of the blue GCs, indicating that the red GCs are more
  concentrated toward the galaxy center than the blue GCs,
  as shown previously (\citealt{forbes04,lee07}).

\subsubsection{Need for an Extended Dark Matter Halo in M60}
\label{dark}

\begin{figure}
\plotone{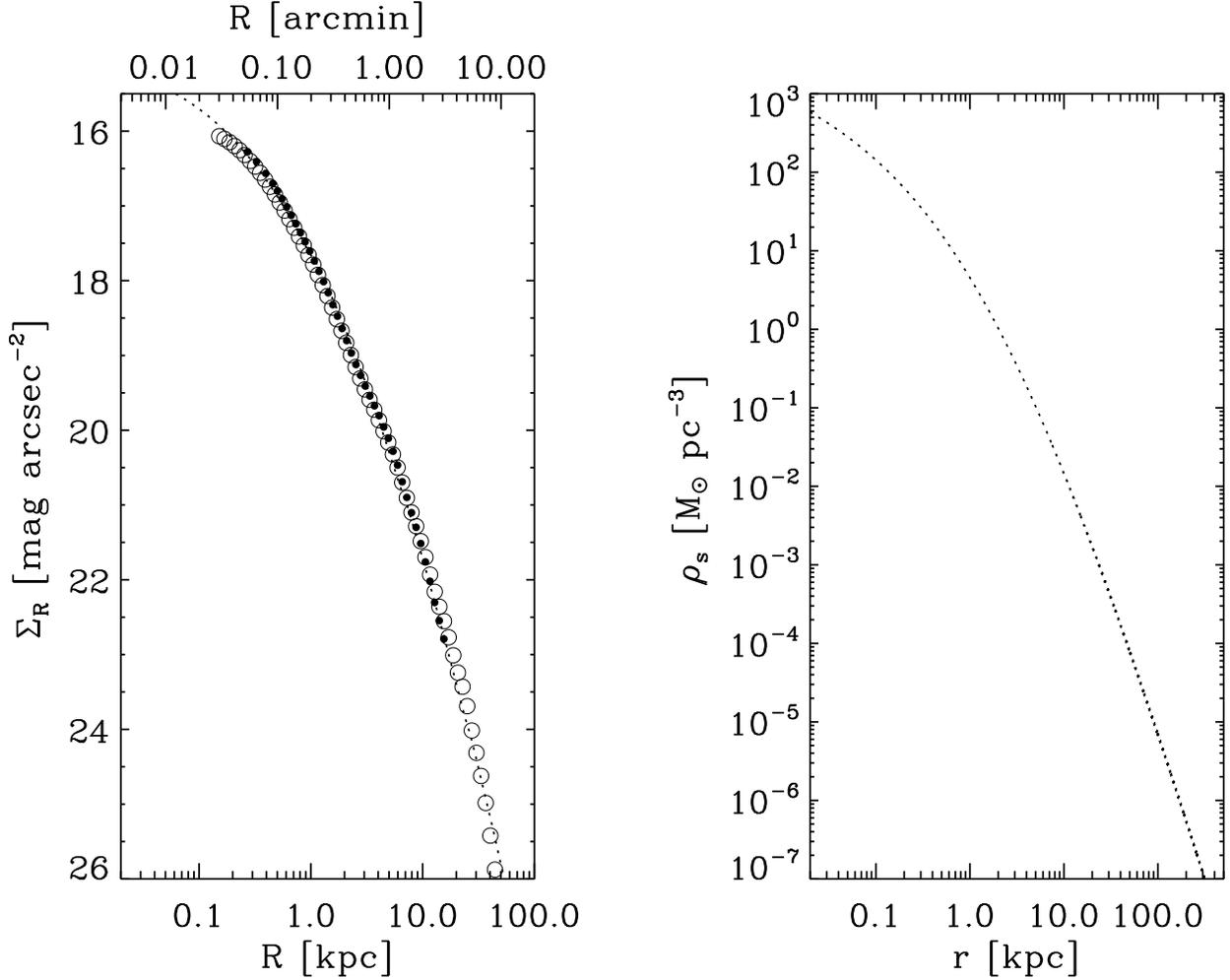} \caption{{\it Left}: $R$-band surface photometry of M60
  derived from KPNO images \citep[open circles]{lee07,kim06} compared to
  that in \citet[filled circles]{pel90}.
Dashed line indicates a projected best fit using eq. (\ref{lumden}).
{\it Right}: Three dimensional stellar mass density profile
  using the best fit model in the left panel with
  a constant $R$-band mass-to-light ratio of $\Upsilon_0=6.0~M_\odot L^{-1}_{R,\odot}$.
\label{fig-surfphot}}
\end{figure}

In the left panel of Figure \ref{fig-surfphot},
  we plotted the surface brightness profile of M60
  derived from our KPNO $T_1$-band images \citep{lee07}
  compared to that in \citet{pel90} for the $R$-band photometry.
We convert $T_1$ photometry of \citet{lee07} to Cousins $R$-band photometry
  using the relation given by \citet{gei96}.
It is seen that two profiles agree well over the radius.

We fit the surface brightness profile derived from the KPNO images \citep{kim06,lee07}
  with the projection of three dimensional luminosity density profile used in \citet{coteetal03},
  which is represented by,
\begin{equation}
j(r) = {{(3-\gamma)(7-2\gamma)}\over{4}}\, {L_{\rm tot}\over{\pi
a^3}}\, \left({r\over{a}}\right)^{-\gamma}\,
\left[1+\Bigl({r\over{a}}\Bigr)^{1/2} \right]^{2(\gamma-4)}\ .
\label{lumden}
\end{equation}

The fit yields the parameters of
  $\gamma=0.32$, $L_{\rm tot}=1.32\times10^{11}~L_{R,\odot}$, and $a=1.48$ kpc,
  and the projected best fit curve is overlaid in Figure \ref{fig-surfphot}.
The fitted model also gives an effective radius of $R_{\rm eff}=1.\arcmin96\simeq9.86$ kpc,
 which is slightly larger than that from a fit ($R_{\rm eff}=1.\arcmin83\simeq9.23$ kpc at $T_1$-band)
 using a de Vaucouleurs law in \citet{lee07}.

In the right panel of Figure \ref{fig-surfphot},
  we show a three dimensional stellar mass density profile, $\rho_s(r)=\Upsilon_0 j(r)$,
  with $R$-band mass-to-light ratio $\Upsilon_0=6.0~M_\odot L^{-1}_{R,\odot}$
  (discussed later in this Section).
Thus we obtain a stellar mass profile of M60, which is represented by
\begin{eqnarray}
M_{\rm s}(r)   & = &\int_{0}^{r}{4{\pi}x^2\,\rho_s(r)}dx  = \Upsilon_0\int_{0}^{r}{4{\pi}x^2\,j(x)}dx \nonumber \\
 & = & {\Upsilon_{0}}L_{\rm tot}\left[{(r/a)^{1/2}
    \over 1+(r/a)^{1/2}}\right]^{2(3-\gamma)}\left[{(7-2\gamma)+(r/a)^{1/2}
    \over 1+(r/a)^{1/2}}\right] \ .
\label{eq-starmass}
\end{eqnarray}

\begin{figure}
\plotone{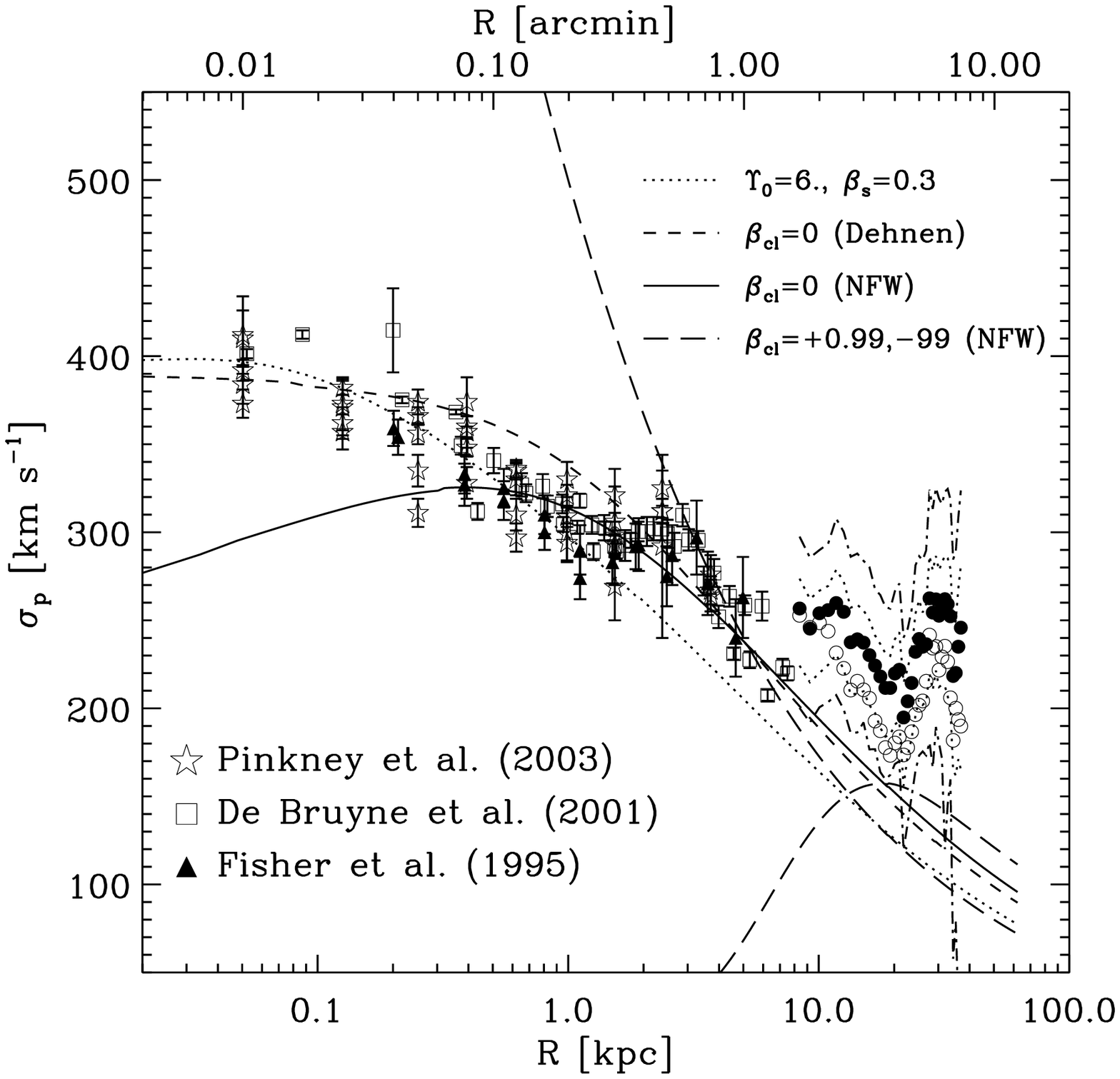} \caption{VDPs for the stars and the GCs.
Stellar VDPs are from \citet[filled triangles]{fisher95},
  \citet[open squares]{debruyne01}, and \citet[open stars]{pinkney03},
  and the GC VDPs shown by filled and open circles with associated, dotted and dot-dashed lines
  are from Fig. \ref{fig-disp}.
Dotted curve represents the stellar VDP calculated using the
stellar mass model
  in Fig. \ref{fig-surfphot} with
  a constant stellar mass-to-light ratio of $\Upsilon_0=6.0~M_\odot L^{-1}_{R,\odot}$
  and a stellar velocity anisotropy of $\beta_{\rm s}=0.3$.
Other lines indicate the VDPs
  calculated using the same stellar mass model as above
  with several GC density profiles and velocity anisotropies:
  Dehnen and $\beta_{\rm cl}=0.0$ (short dashed line),
  NFW and $\beta_{\rm cl}=0.0$ (solid line),
  NFW and $\beta_{\rm cl}=+0.99,-99$ (long dashed lines).
\label{fig-stardisp}}
\end{figure}

We used this stellar mass profile to determine the velocity
  anisotropy for the M60 stellar system and to test the existence of
  an extended dark matter halo.
If we take $M_{\rm tot}(r)=M_{\rm s}(r)$
  and substitute $n_{\rm cl}(r)$ by $\rho_s(r)\propto j(r)$,
  then we can compute the intrinsic radial VDP of the stars
  through the Jeans equation
  by assuming several $R$-band mass-to-light ratios ($\Upsilon_0$) and
  velocity anisotropies of the stellar system ($\beta_{\rm s}(r)$).
We therefore predict the projected VDPs for the stellar system
  from the intrinsic radial VDP calculated using equation (\ref{eq6}).
In Figure \ref{fig-stardisp}, we show the projected VDPs
  calculated with $\Upsilon_0=6.0~M_\odot L^{-1}_{R,\odot}$ and $\beta_{\rm s}(r)=+0.6$ (radially biased),
  which is the best fit curve for the stellar kinematic data of
  \citet{fisher95}, \citet{debruyne01}, and \citet{pinkney03}.
This $R$-band mass-to-light ratio and radially biased velocity anisotropy
  for the stellar system is similarly found in the M49 stellar system
  with $\Upsilon_0=5.9~M_\odot L^{-1}_{R,\odot}$ and $\beta_{\rm s}(r)=+0.3$
  (\citealt{coteetal03} and references therein).

For the comparison, we also present the projected VDPs calculated
  using the same stellar mass profile as above,
  but for the GC number density profile $n_{\rm cl}(r)$
  and for $\beta_{\rm cl}(r)=+$0.99 (radially biased, upper long dashed lines),
  $-$99 (tangentially biased, lower long dashed lines),
  0.0 (isotropic, solid line for the NFW profile and short dashed line for the Dehnen profile).
Interestingly, none of these models can account for
  the observed VDPs for the GCs at $R>7$ kpc, indicating that
  mass-to-light ratio is not constant over the galactocentric
  distance, but should be increased as the distance increases.
This means that there exists an extended dark mater halo in M60.
This result is consistent with the previous findings of
  an extended dark mater halo in M60
  from a radial profile of an increasing mass-to-light ratio
  in $K$-band \citep{hum06} and in $V$-band \citep{bri06}
  at $7\lesssim R\lesssim22$ kpc.

\subsubsection{X-ray Mass Profile}
The total gravitating mass profile is determined using a gas
temperature profile and a density distribution obtained from X-ray
observational data on the assumption of hydrostatic equilibrium
(neglecting magnetic pressure term):

\begin{equation}
M_{\mathrm {tot}}(r) = - {{k T(r) r} \over {G \mu m_p}} \left(
{d\log\rho(r) \over d\log r} + {d\log T(r) \over d\log r} \right),
\label{eqmass}
\end{equation}

where $k$ is the Boltzmann constant, $G$ is the gravitational
constant, $T(r)$ is a gas temperature at radius $r$, $\mu$ is
the mean molecular weight (taken as 0.63 in this study), $m_p$ is the
proton mass, and $\rho(r)$ is a gas density at radius $r$.
For the case of M60, \citet{brighenti97} derived the total mass
  profile using {\it Einstein} HRI observational data of \citet{trinchieri86} and {\it ROSAT} PSPC observational data
  of \citet{trinchieri97}, and \citet{hum06} derived the total mass profile using {\it Chandra} observational data.
In addition, \citet{randall06} reported the density and the temperature distribution of M60 using {\it XMM-Newton} data.
Since the derived VDP is sensitive to the mass profile,
we estimate mass profiles using these different X-ray data set.

\begin{figure}
\plotone{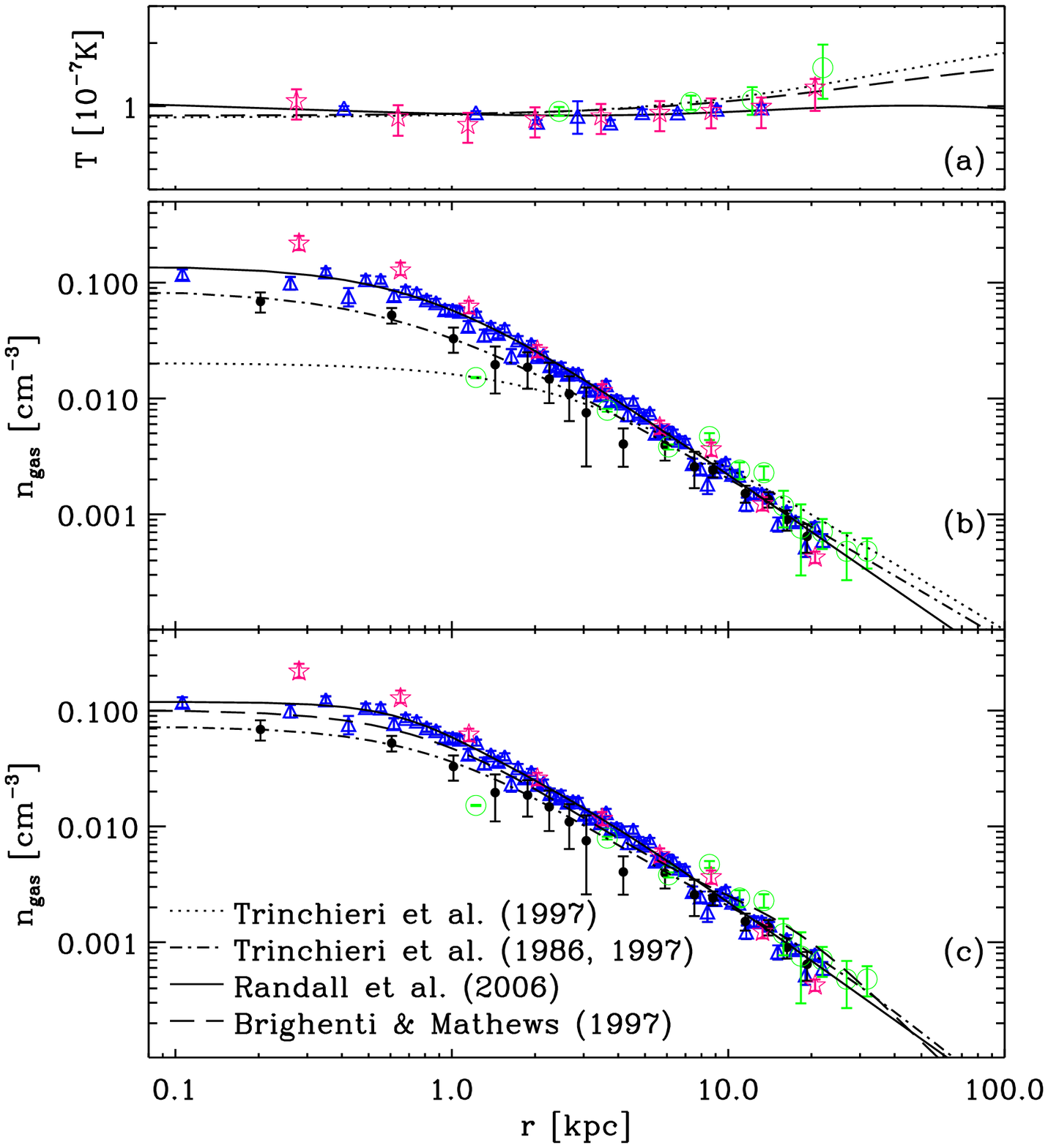} \caption{Radial profiles of the gas temperature (a),
  the gas number density with one component fit (b),
  and the gas number density with two component fit (c).
The different symbols represent the data from different literature:
  \citet[open circles]{trinchieri97}, \citet[filled circles]{trinchieri86},
  \citet[open triangles]{randall06}, and \citet[open stars]{hum06}.
The dotted, dot-dashed and solid lines in each panel indicate the
best fit
  curves using the annotated data, while the long dashed lines in (a)
  and (c) denote the profiles derived by \citet{brighenti97}.
\label{fig-xrayfit}}
\end{figure}

In Figure \ref{fig-xrayfit}, we display the deprojected profiles of the gas temperature (top panel) and
  the gas number densities (middle and bottom panels) found by
  \citet[filled circles]{trinchieri86}, \citet[open circles]{trinchieri97},
  \citet[open triangles]{randall06}, and \citet[open stars]{hum06}
  with measured errors.
Having done this, we fit the temperature and
  gas number density data of each reference using the
  temperature distribution of
  $T(r)=2T_m[r_m/(r+r_{ot})+(r/r_m)^q]^{-1}$, and the density
  distribution of $n(r) = \Sigma n_i (r)$, where $n_i(r) = n_0(i)
  [1+\left\{r/r_0(i) \right\}^{p(i)}]^{-1}$.
$T_m$, $r_m$, $r_{ot}$, $q$, $n_0(i)$, $r_0(i)$, and $p(i)$ are parameters of the fit.
We fit the gas number density profile of each data for two cases of $i\le1$ (one
  component fit) and $i\le2$ (two component fit). The data in use
  are annotated with the associated lines. The results for the fit are
  summarized in Table \ref{tab-xrayfit}.

\begin{figure}
\plotone{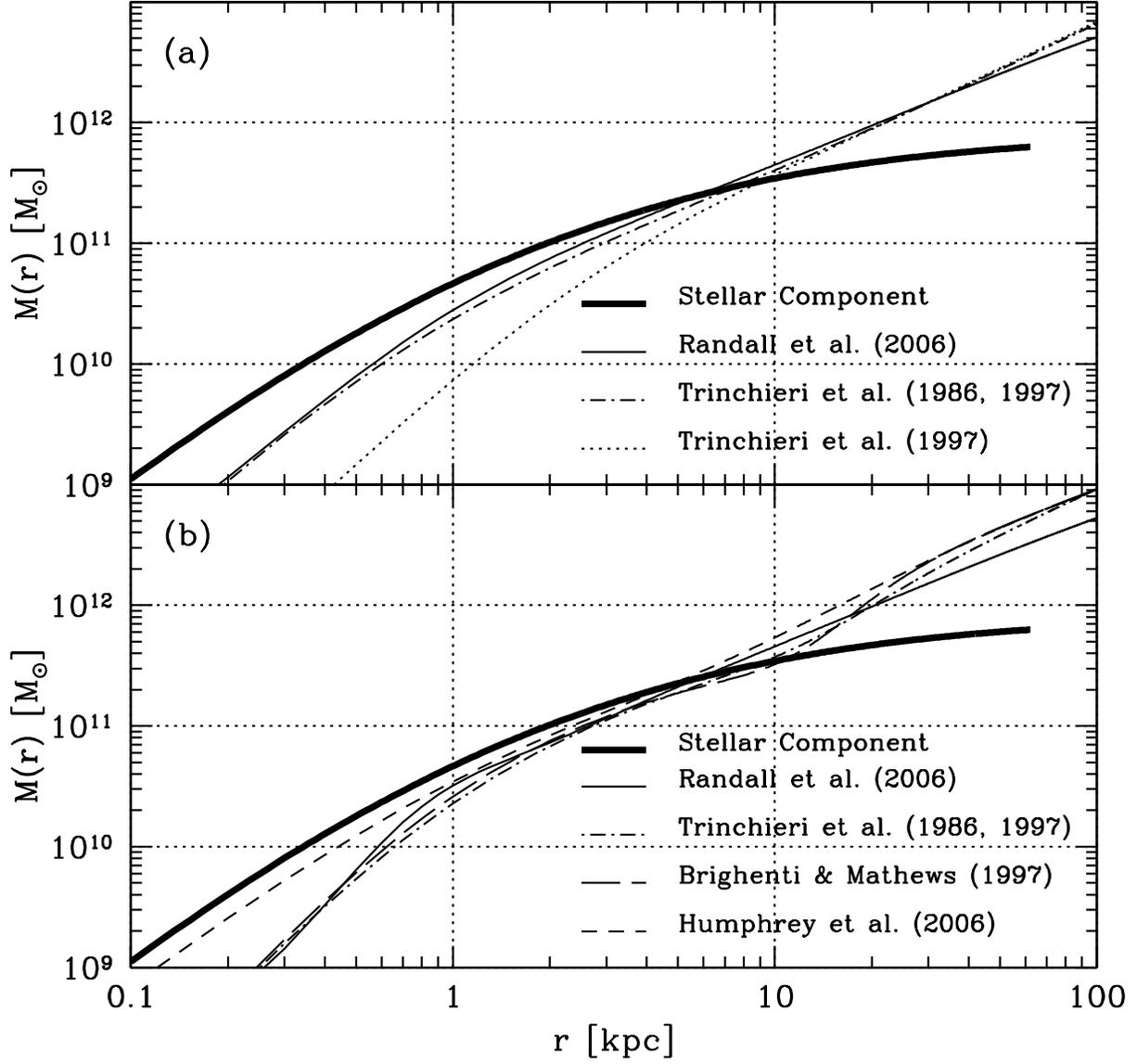} \caption{Total mass profiles of M60 derived (a)
  using the gas number density profile with one component fit,
  and (b) using the gas number density profile with two component fit.
The solid, dot-dashed and dotted lines indicate the mass profiles
  derived in this study using the annotated data, while the long dashed line and the short dashed line in (b)
  denote the M60 mass profiles derived by \citet{brighenti97} and \citet{hum06}, respectively.
A heavy solid line represents the stellar mass profile
 using a constant $R$-band mass-to-light ratio of $\Upsilon_0=6.0~M_\odot L^{-1}_{R,\odot}$
 derived in \S \ref{dark}.
\label{fig-mass}}
\end{figure}

The resulting mass profiles of M60, $M(r)$, for several temperature and
  gas number density profiles are presented in Figure \ref{fig-mass}.
The upper panel shows the mass profiles using one
  component fit of the gas number density for various data set.
The lower panel shows those using two components of the
  gas number density for various data set and using the mass profile
  derived by \citet{hum06} in comparison. It appears that most
  mass profiles agree well.
However, the mass profile derived using the
  temperature and the gas number density data (with one
  component fit) of \citet[the dotted line in the upper
  panel]{trinchieri97} deviates from the other profiles in the inner region ($r\leq 10$ kpc).
In addition, the mass profile derived by \citet[the short dashed line in the lower
  panel]{hum06} deviates from the other profiles in the region of $r\leq 1$ kpc.
Comparing X-ray mass profiles with the stellar mass profile (heavy
  solid line) determined in \S \ref{dark},
  we find that most X-ray mass profiles deviate significantly from the stellar mass profiles
  in the inner region at $r<2$ kpc, indicating that X-ray mass profiles
  are not reliable at the very small radius because of
  angular resolution limit and non-equilibrium energetics.
In the intermediate region at $2<r<10$ kpc,
  the X-ray mass profiles derived using two component fit of the gas
  number density (lower panel) agree to the stellar mass profiles,
  while those derived using one component fit of the gas
  number density (upper panel) do not.
In the outer region at $r>10$ kpc, all X-ray mass profiles are
  larger than the stellar mass profiles, confirming the need of
  an extended dark matter halo as shown in \S \ref{dark}.
Since the discrepancy between X-ray and stellar mass profiles
  is significant only in the inner region at $r<2$ kpc, where
  there are no GCs, and X-ray mass profiles that account for the
  dark matter halo are similar to or
  larger than the stellar mass profile at $r>2$ kpc,
  we conclude that the X-ray mass profiles are good enough to determine
  the velocity anisotropy of the M60 GC system for the following analysis.

\subsubsection{Determination of the Velocity Anisotropy}
We determine the velocity anisotropy of GCs as follows:
  (1) With the GC number density profile ($n_{\rm cl}(r)$)
  of the entire, blue, and red GCs and the mass profile ($M_{\rm tot} (r)$) in hand,
  assuming the velocity anisotropy ($\beta_{\rm cl}(r)$) in prior,
  we derive the theoretical projected VDP ($\sigma_p(R)$) and theoretical
  projected aperture VDP ($\sigma_{ap}(\le R)$) using the Jeans equation;
  (2) From the comparison of these calculated VDPs with measured VDPs,
  we determine the velocity anisotropy of GCs.

We begin by deriving the theoretical projected VDPs.
The equation (\ref{eq4}), spherical Jeans equation can
be solved for the radial component of velocity dispersion,
$\sigma_r(r)$:
\begin{equation}
\sigma_r^2(r)={1\over{n_{\rm cl}(r)}}\, \exp\left( -\int
{{2\beta_{\rm cl}}\over{r}}\,dr \right)\, \left[ \int_r^{\infty}
n_{\rm cl}\,{{GM_{\rm tot}}\over{x^2}}\, \exp\left( \int
{{2\beta_{\rm cl}}\over{x}}\,dx \right)\, dx \right]\ .
\label{eq5}
\end{equation}
Then the projected VDP, $\sigma_p(R)$ can
be derived by
\begin{equation}
\sigma_p^2(R) = {2\over{N_{\rm cl}(R)}}\, \int_R^{\infty} n_{\rm
cl} \sigma_r^2(r) \left(1 - \beta_{\rm cl}\,{{R^2}\over{r^2}}
\right)\, {{r\,dr}\over{\sqrt{r^2 - R^2}}}, \label{eq6}
\end{equation}
where $R$ is the projected galactocentric distance and the surface density profile, $N_{\rm cl}(R)$, is a
  projection of the three dimensional density profile $n_{\rm cl}(r)$.
The projected aperture VDP, $\sigma_{\rm ap}(\le R)$, which is the
velocity dispersion of all objects interior to a given projected
radial distance $R$, can be computed by
\begin{equation}
\sigma_{\rm ap}^2(\le R) =
  \left[ \int_{R_{\rm min}}^R N_{\rm cl}(R^\prime) \sigma_p^2(R^\prime)\,
         R^\prime\,dR^\prime \right]\,
  \left[\int_{R_{\rm min}}^R N_{\rm cl}(R^\prime)\,R^\prime\,dR^\prime
         \right]^{-1}\ ,
\label{eq8}
\end{equation}
where $R_{min}$ is the projected galactocentric distance of the
innermost data point in the GC sample ($R_{min}=2.7$ kpc in this
study).

\begin{figure}
\plotone{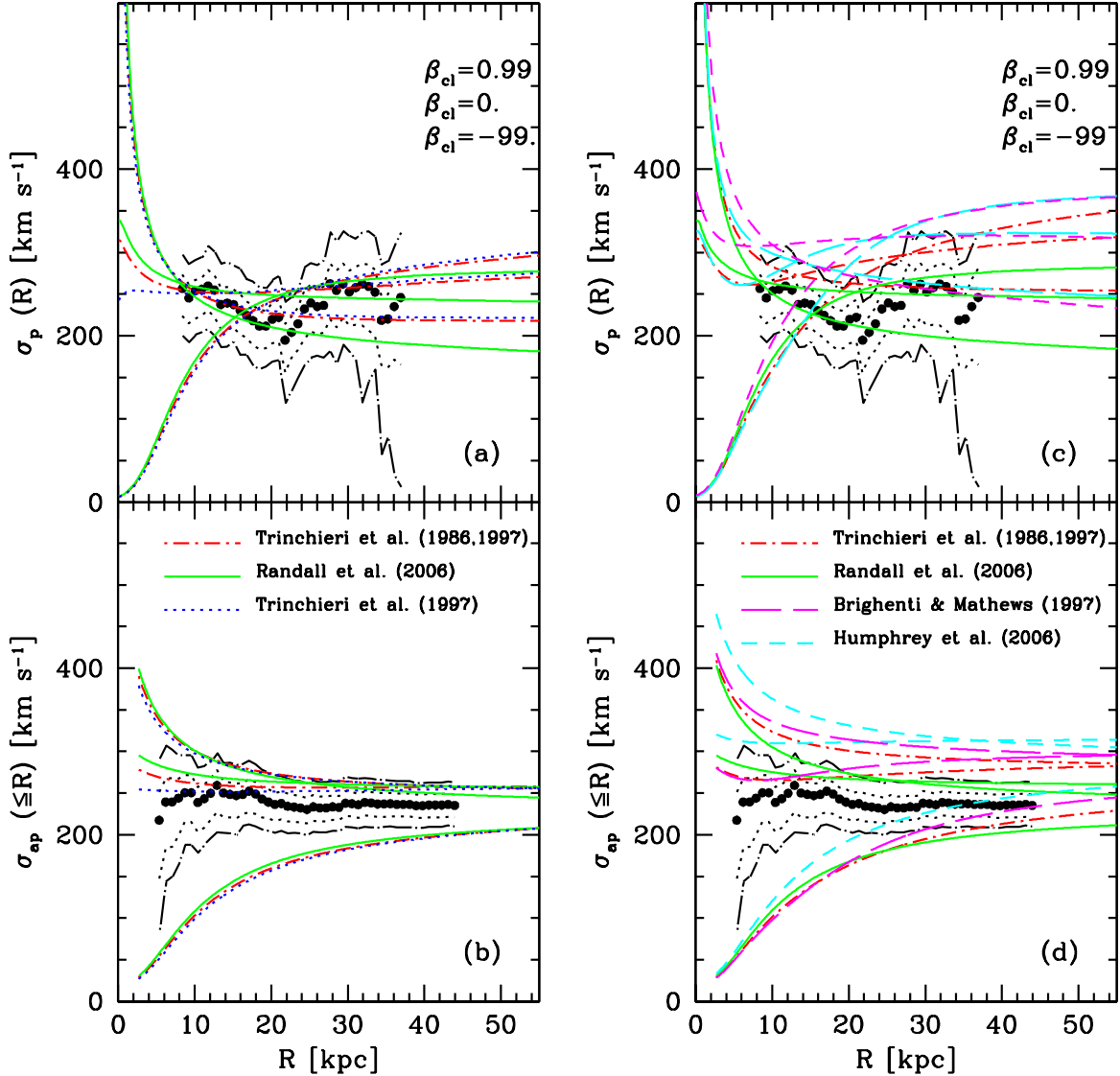} \caption{VDPs (a, c) and
  the aperture VDPs (b, d) for the entire GCs.
Filled circles in (a, c) represent the measured VDP shown in Fig. \ref{fig-disp},
  and those in (b, d) denote the measured aperture VDP.
Associated, dotted and dot-dashed lines represent 68\% and
  95\% confidence intervals on the calculation of velocity
  dispersion, respectively.
Three smoothly curved lines, from a radially biased
  velocity anisotropy to the tangentially biased velocity anisotropy
  (from top to bottom, $\beta_{\rm cl}$= 0.99, 0, and $-$99),
  represent the VDPs calculated using
  the GC density profile of Dehnen with one component fit of the
  gas number density profile (a, b) and with two component fit
  of the gas number density profile (c, d).
\label{fig-isoedh}}
\end{figure}

\begin{figure}
\plotone{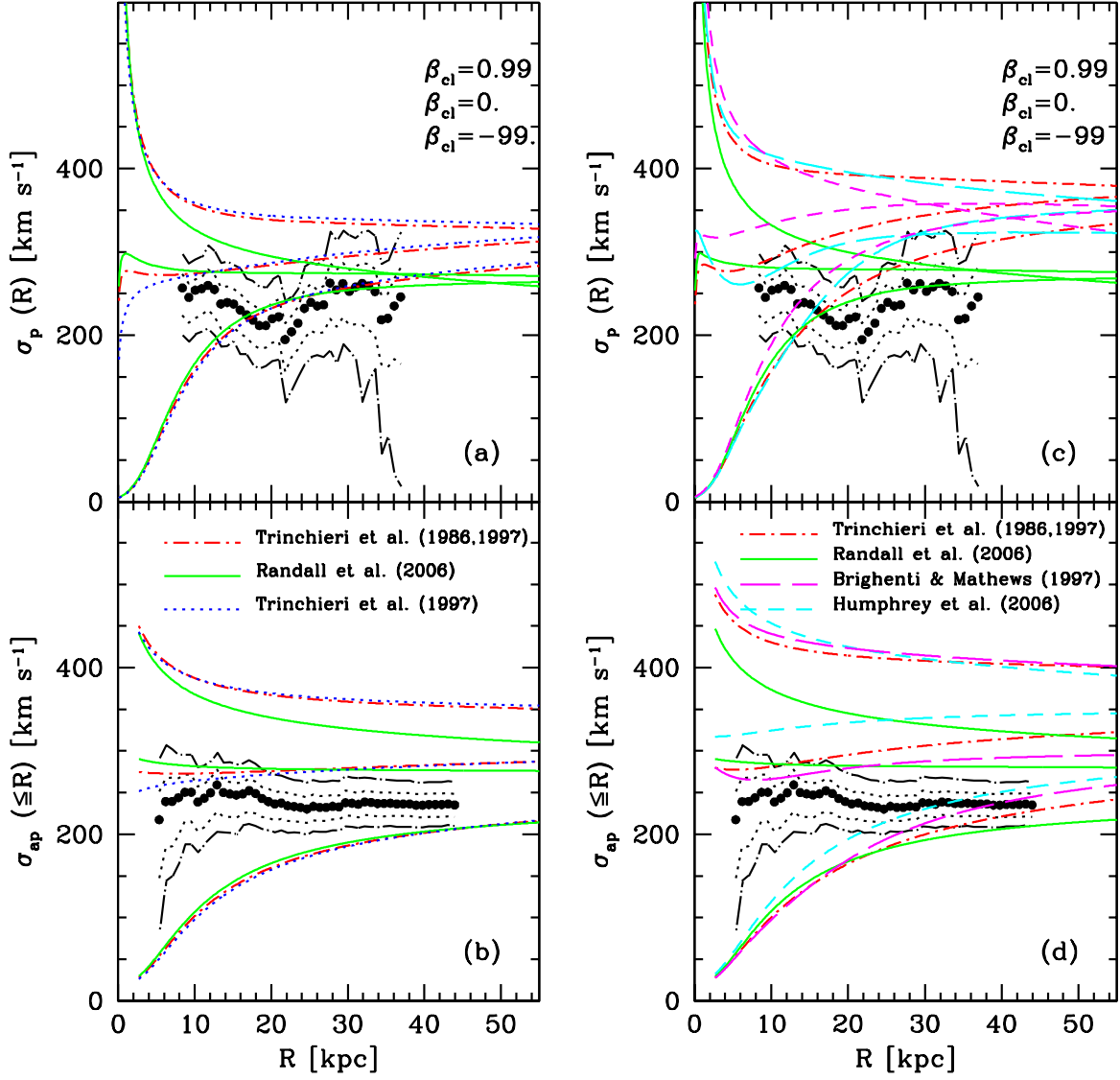} \caption{Same as Figure \ref{fig-isoedh}, but
for the VDPs calculated using the GC density profile of NFW.
\label{fig-isoenfw}}
\end{figure}

We present the measured VDP in comparison with the
  VDPs calculated by assuming several velocity anisotropies
  in Figures \ref{fig-isoedh} and \ref{fig-isoenfw}.
Figure \ref{fig-isoedh} shows the VDPs calculated using
  the Dehnen profile for the GC number density, while Figure
  \ref{fig-isoenfw} shows those using the NFW profile for the GC
  number density.
The upper panels show the projected VDPs, and the lower panels
show the projected aperture VDPs. The measured dispersion data
taken from Figure \ref{fig-disp} are shown by filled circles along
with their confidence intervals. The projected aperture VDPs in
  the lower panels are plotted in the similar fashion to the case
  of the upper panel.
The calculated VDPs in the left panels are obtained
  with the mass profiles derived using one component fit of the gas
  number density, while those in the right panels are obtained with
  the mass profiles derived using two component fit of the gas number density.
Although it is difficult to distinguish the velocity anisotropy
  clearly for nearly all radial distances in the upper panel (the bottom panels show a more stable result),
  it appears that the entire GC system
  does not have an isotropic velocity ellipsoid ($\beta_{\rm cl}=0$), but have a modest
  tangentially biased velocity ellipsoid ($\beta_{\rm cl}<0$) for any mass profile in Figure
  \ref{fig-isoedh} based on the Dehnen profile for the GC number density.
A similar result can be found in Figure~\ref{fig-isoenfw} based on
  the NFW profile for the GC number density.
\citet{bri06} reported that the orbits of M60 GCs are close to isotropic within
  100$\arcsec$ ($\sim$8.4 kpc) and becomes tangentially biased
  beyond 100$\arcsec$.
Our results appear to be consistent with those of \citet{bri06} as a whole,
  although the signature of an isotropic orbit within 100$\arcsec$ is weaker in this study.
However, we extend the radial coverage ($\sim$21 kpc) of
\citet{bri06} out to $\sim$40 kpc in this study.

\begin{figure}
\plotone{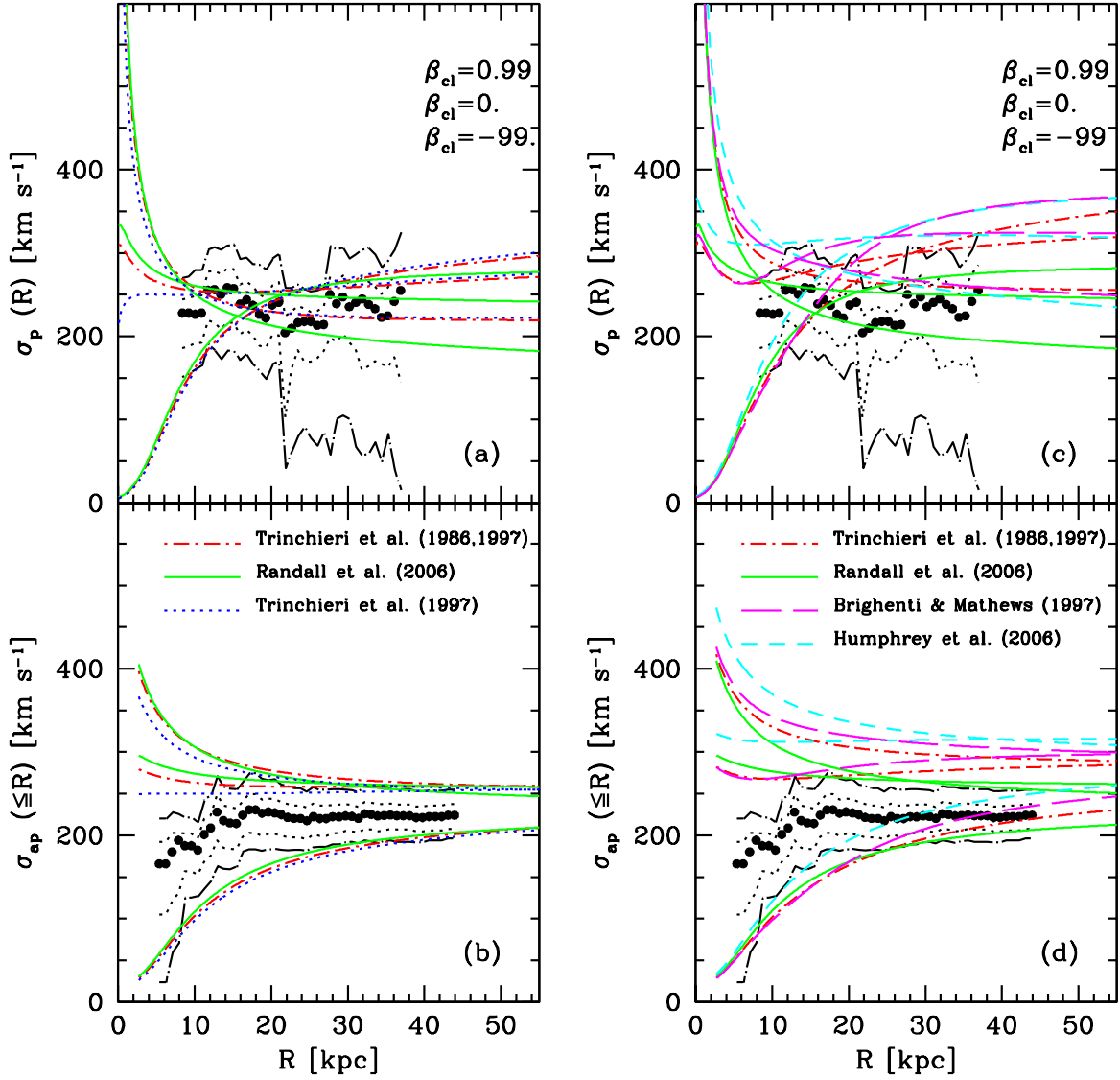} \caption{Same as Figure \ref{fig-isoedh}, but
for the blue GCs. \label{fig-isobdh}}
\end{figure}

\begin{figure}
\plotone{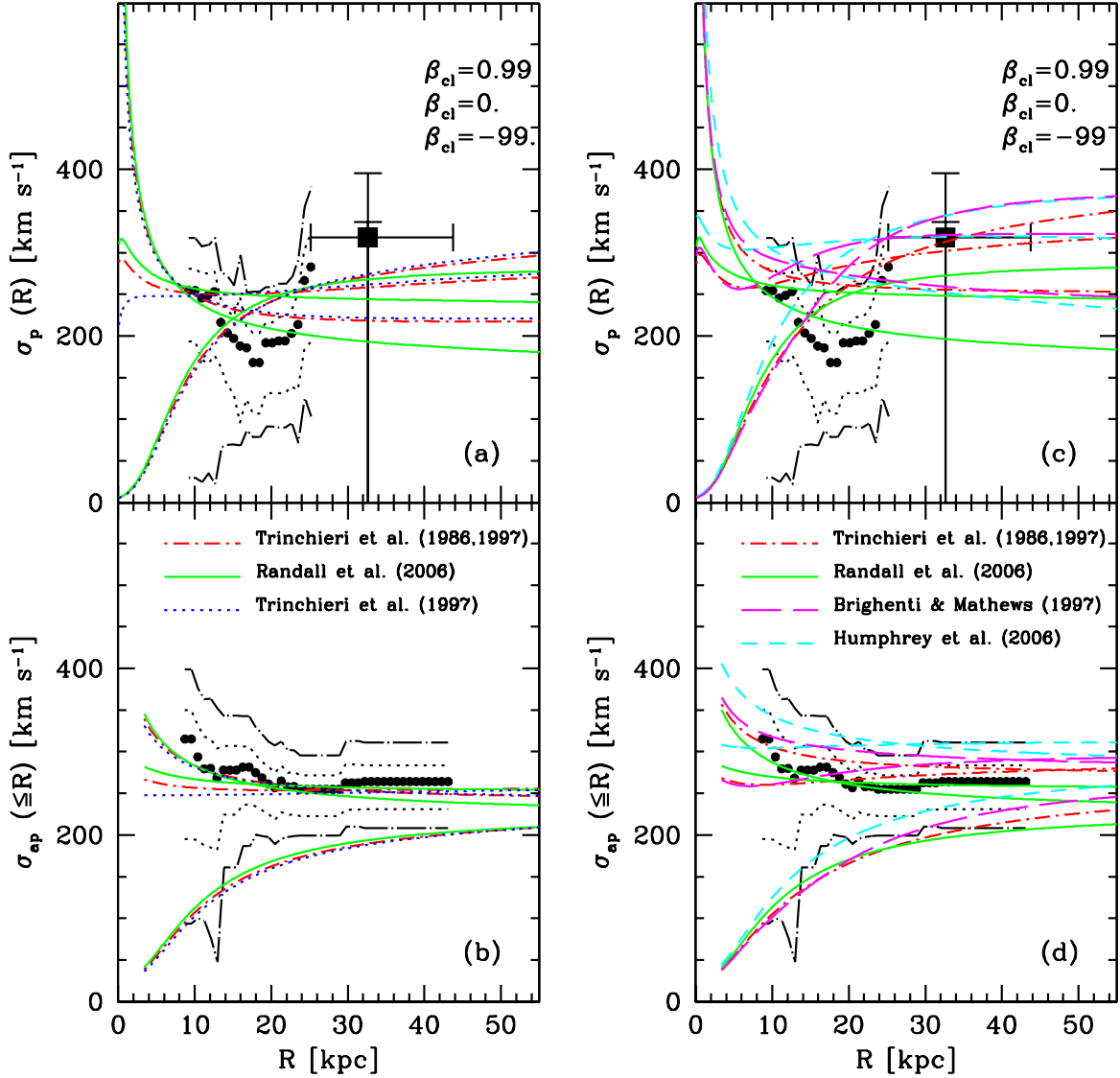} \caption{Same as Figure~\ref{fig-isoedh}, but for the red GCs.
In the top panels, large filled square represents the measured velocity dispersion about the mean GC velocity ($\sigma_p$)
beyond $R\simeq25.1$ kpc. \label{fig-isordh}}
\end{figure}

In Figures~\ref{fig-isobdh} and \ref{fig-isordh},
  we show a similar analysis for the blue and red GCs, respectively.
We present the results using the Dehnen profile of GC number density for
the blue and red GCs. The results using the NFW profile are not different from
those in Figures~\ref{fig-isobdh} and \ref{fig-isordh}.
Although it is not easy to draw a strong conclusion
  due to the small number statistics and complex mass profiles,
  we note a difference of velocity ellipsoids between the blue and red GCs in the projected
  aperture VDP (the bottom panels in Figure~\ref{fig-isobdh} and \ref{fig-isordh}).
It appears that the blue GC system has a tangentially biased velocity ellipsoid with $\beta_{\rm cl}<0$,
while the red GC system has a radially biased or an isotropic velocity ellipsoid.

\section{Discussion}
\subsection{Comparison with the GC Systems in Other gEs}\label{comp_ge}

To date there are five giant elliptical
galaxies except for M60 that the kinematics of their GC systems was studied:
  M87 \citep{cohen97,kissler98,cote01}, M49 \citep{zepf00,coteetal03},
  NGC 1399 \citep{kissleretal98,minniti98,kissler99,richtler04},
  NGC 5128 \citep{peng04,wood07}, and NGC 4636 \citep{schuberth06}.

For the comparison of the kinematic properties of GCs in those
gEs, we analyze the velocity data of 276 GCs in M87
\citep{cote01}, 263 GCs in M49 \citep{coteetal03}, 435 GCs in NGC
1399 \citep{richtler04}, 210 GCs in NGC 5128 \citep{peng04}, and
172 GCs in NGC 4636 \citep{schuberth06} using the similar method
adopted in this study. M87, M49, and NGC 4636 as well as M60 are
gEs in the Virgo cluster, and NGC 1399 is a gE in the Fornax
cluster at the similar distance to that of the Virgo cluster. NGC
5128 is in the Centaurus group, being the nearest gE. Basic
photometric properties of these galaxies are listed in
Table~\ref{tab-gesample}. The column (1), (2), and (3) give the
name of galaxy, the absolute magnitude in the $V$ band, and the
systemic radial velocity, respectively. The effective radius,
the ellipticity, the position angle of photometric minor axis in
degrees east of north, and the distance are shown in the column
(4), (5), (6), and (7), respectively. The resulting global
kinematics for the entire, blue and red GCs of each galaxy are
presented in Table~\ref{tab-gekin}: the mean radial velocity,
velocity dispersion about the mean radial velocity, rotation
axis, rotation amplitude, rotation-corrected velocity dispersion,
and the absolute value of the ratio of the rotation amplitude to
the velocity dispersion about the best fit sine curve for each GC
subsample. We divide the GCs in each gE into the blue and red GCs
using the colors that were used in the associated reference,
excluding the GCs without color information, for the kinematic
analysis. In addition, we estimate the rotation amplitude and the
position angle of rotation axis for the GC system of NGC 5128, by
fitting the mean velocity of GCs lying in a fixed width (60
deg) of position angle for a stable computation. The velocity
dispersions derived in this study agree well with those derived in
the associated reference. The velocity dispersions of GCs in M87
($\sigma_p=414^{+ 15}_{-18}$ km s$^{-1}$) and NGC 1399
($\sigma_p=323^{+ 11}_{-13}$ km s$^{-1}$) are found to be larger
than and comparable to that of GCs in M49 ($\sigma_p=322^{+
14}_{-17}$ km s$^{-1}$), respectively, although M87 and NGC 1399
are fainter than M49. This is due to the fact that M87 and NGC
1399 are located in the center of a galaxy cluster, while M49 is
not. NGC 5128 shows the smallest velocity dispersion
($\sigma_p=129^{+  5}_{ -7}$ km s$^{-1}$) among our sample
galaxies, implying a relatively smaller mass in spite of its high
luminosity in the $V$ band ($M_V=-21.7$).

Table~\ref{tab-gesum} lists several notable features
  based on the global kinematics presented in Table~\ref{tab-gekin}.
The strength of rotation is defined in the column (4):
  strong for $\Omega R$/${\sigma}_{p,r}>$ 0.4, modest for 0.4 $\ge \Omega R$/${\sigma}_{p,r}>$ 0.2,
  and weak for $\Omega R$/${\sigma}_{p,r} \leq 0.2$. The rotation axis in the column (5)
is assigned if the difference between the position angle of
rotation axis and that of photometric major/minor axis is less
than 30$^\circ$. The result of velocity anisotropy for each GC
system is taken from the literature (see column (6)). It appears
that the rotation-corrected velocity dispersion,
  $\sigma_{p,r}$ of the blue GCs is similar to or larger than that of
  the red GCs except for M60. This implies that the blue GC system is
  dynamically hotter than the red GC system in most gEs.
However, the case of rotation ($\Omega R$/${\sigma}_{p,r}$) is not simple.
Both of the blue and red GCs in M60 and M87 show a strong rotation, while
  those in NGC 1399 show a weak rotation.
The red GCs in NGC 4636 and NGC 5128 show a slightly stronger rotation than the blue GCs.
However, the red GCs in M49 show a weak rotation, while the blue GCs show a modest rotation,
  which is consistent with the prediction of the merger formation model \citep{ashman92}.
In addition, if we consider the position angle
  of rotation axis together, the story becomes more complicated.
In velocity anisotropy, it appears that the entire sample of the
GC system in three gEs (M87, M49, and NGC 1399) has an isotropic
velocity ellipsoid and two subsamples
  have different velocity ellipsoids in two gEs (M60 and M87).
It is needed to determine the velocity anisotropy of the GC
systems in more gEs
  including NGC 5128 and NGC 4636.

\begin{figure}
\plotone{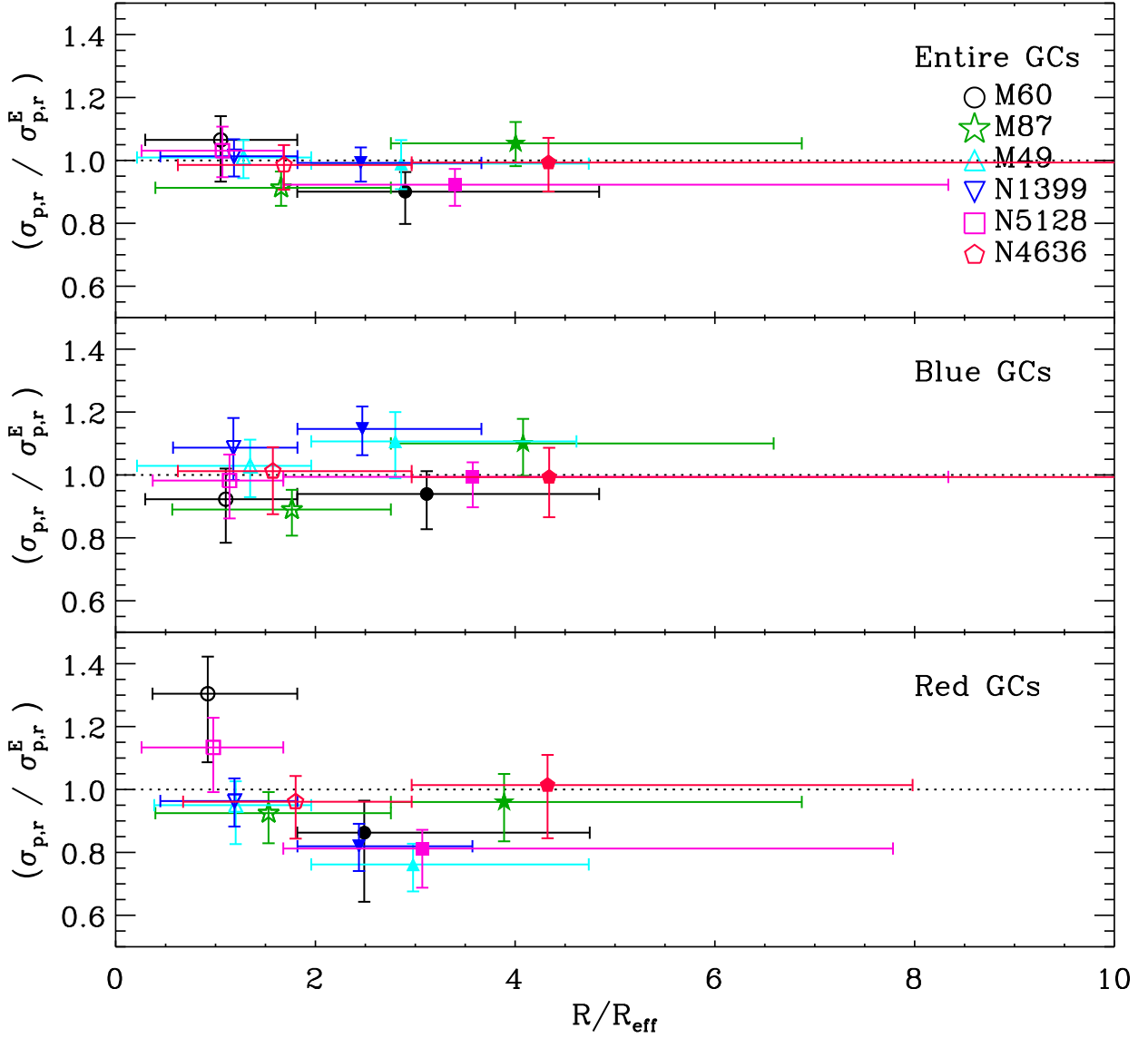} \caption{Rotation-corrected velocity dispersions
in gEs versus the projected galactocentric distances for the
entire (top panel), blue (middle panel), and red (bottom panel)
GCs. The rotation-corrected velocity dispersion is normalized with
respect to that of the entire GCs in each gE. The projected
galactocentric distance is normalized with respect to the
effective radius of each gE. Open symbols indicate the dispersions
in the inner region of each gE, while filled symbols those in the
outer region. \label{fig-gesigresi}}
\end{figure}

\begin{figure}
\plotone{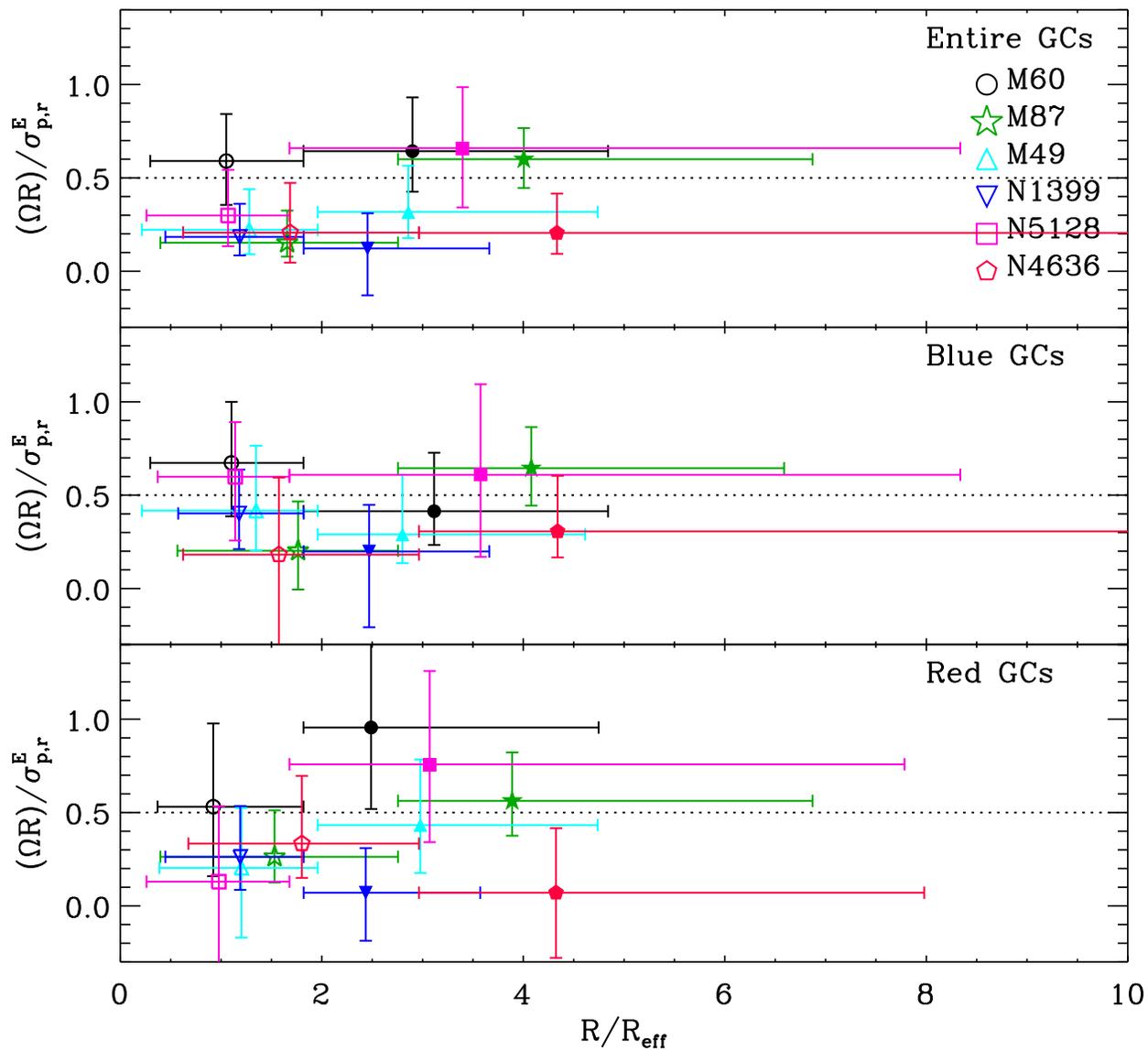} \caption{Absolute values of the ratio of rotation amplitude
to velocity dispersion versus the projected galactocentric distances for the entire (top panel),
blue (middle panel), and red (bottom panel) GCs. Open symbols indicate
the ratios in the inner region of each gE, while filled symbols those in the outer region. \label{fig-gerot}}
\end{figure}

To examine the general kinematic properties of the GC systems in
gEs, we plot the rotation-corrected velocity dispersions in gEs
against the projected galactocentric distances in Figure
\ref{fig-gesigresi}. The rotation-corrected velocity dispersion is
normalized by that of the entire GCs in each gE. The projected
galactocentric distance is normalized by the effective radius of
each gE. The entire and blue GCs do not show significant change of
the velocity dispersion over the whole region of a galaxy.
However, the velocity dispersion of the red GCs in the inner
region ($R<2R_{\rm eff}$) is, in the mean, marginally larger than that
in the outer region ($R>2R_{\rm eff}$). This implies that red GC
system in the inner region may be dynamically hotter than that in
the outer region. In Figure \ref{fig-gerot}, the absolute value of
the ratio of the rotation amplitude to the velocity
  dispersion is plotted as a function of the projected
  galactocentric distance.
For the red GCs, the ratio of the rotation
  amplitude to the velocity dispersion in the outer region appears to be marginally larger
  than that in the inner region if we neglect the point of NGC 4636 (filled pentagon),
  while it does not change with the distance for the entire and blue GCs.

In summary, differences in kinematic properties among GC
  subsamples appear to exist, although it depends on the galaxy. The blue GC
system appears to be dynamically similar to or hotter than the red GC system,
while the rotation of the GC system is not negligible. The entire
sample of each GC system appears to have an isotropic velocity ellipsoid,
while the subsamples do not have uniform velocity ellipsoids. The
kinematic properties of M60 and M87 are similar except for the
velocity dispersion of subsamples and the velocity anisotropy of the entire
GC system, while other galaxies have
diverse kinematic properties. For the red GCs, the velocity
dispersion of the inner region is marginally larger than that of the outer
region, while the rotation of the outer region appears to be more significant
than that of the inner region.

\subsection{Formation Models of Globular Clusters}

The kinematic properties of the GC system in M60 compared to those
in other gEs are useful to test the kinematic predictions of
formation models of the GCs in gE. A summary of several model
descriptions and predictions can be found in several literature
\citep{rhode01,lee03,richtler04,west04,brodie06}. Classical
formation models can be divided into four broad categories: the
monolithic collapse model, the major merger model, the multiphase
dissipational collapse model, and the dissipationless accretion
model.

The monolithic collapse model describes that an elliptical galaxy
and its GCs are formed through the collapse of an isolated massive
gas cloud or protogalaxy at high redshift
\citep{larson75,carlberg84,arimoto87}. In this model, the color
distribution of GCs shows a smooth shape with a single peak, and
the rotation of GCs can be generated by tidal torques from
companions \citep{peebles69}. Although this model can explain some
observational properties of elliptical galaxies successfully (see
\citealt{chiosi02}), the bimodal color distribution of the GC
system in many gEs (e.g., \citealt{lee03,peng06}) make it hard to
accept this model. In addition, the strong rotation of the GC
system as seen in M60 and M87 is not expected from the collapse of
a single, isolated protogalactic cloud.
Moreover, several GC systems in gEs show a globally isotropic velocity
  ellipsoid that needs some kinds of relaxation processes that the
  monolithic collapse model can not account for.

The major merger model suggests that elliptical galaxies are formed by
a merger of two or more disk galaxies
\citep{toomre77,ashman92,zepf00}.
In this model, younger, spatially concentrated, red GCs are formed during the merger, while
spatially extended, blue GCs come from the halos of the disk
galaxies (e.g., \citealt{bekki02}). As a result, the color distribution of GCs is expected to be bimodal.
This model predicts that the newly formed red GCs show little
rotation compared to the blue GCs since the angular momentum would be
transported to the outer region during the merging process. This
model has received particular attention since it could explain
several photometric properties of the GC system in gEs and little
rotation of the red GC system in M49 \citep{zepf00}.
However, contrary to the case of M49, the red GC system in M60, M87,
NGC 5128, and NGC 4636 show a strong or modest rotation.
From the simulation of dissipationless major mergers of spiral galaxies, \citet{bekki05}
  found that both pre-existing metal-poor clusters (MPCs) and metal-rich clusters (MRCs)
  obtain significant amounts of rotation beyond $\sim$ 10 kpc,
  regardless of the orbital configuration of the merging galaxies.
However, both the blue and red GC systems in NGC 1399 show weak rotation,
  which is not consistent with the result of \citet{bekki05}.
Therefore, it is needed to explain what makes the complex rotational properties of the
  GC systems in gEs.
In the dissipationless major merger model, 
  most GCs are formed before a last major merger event. 
There are several observational results based on the spectroscopy of the GCs in gEs 
  consistent with this: the blue GCs and red GCs show a small age difference and are both old : 
  M60 \citep{pierce06}, 
  M87 \citep{coh98}, 
  M49 \citep{bea00,coh03}, 
  and NGC 1399 \citep{kissleretal98,for01}. 
Then, what we observe today is the GC system that was affected by the orbital mixing 
  caused by the last major merger event, 
  and the observed kinematics can not reflect the circumstance when the GC system was formed. 
Therefore, it is noted that it is difficult to trace the orbital history of the GC system 
  from the observed kinematics \citep{kissler98}, 
  and to test the validity of the dissipationless merger scenario.

\citet{forbes97} proposed the multiphase dissipational collapse
model that ellipticals form their GCs in distinct star formation
phases through a dissipational collapse. In addition, there is a
capture of additional GCs by tidal effects from neighboring
galaxies or the accretion of dwarf galaxies. Since the blue GCs
are formed in the first star formation phase and the red GCs are
formed in the subsequent star formation phase after the gas in the
galaxy is self-enriched, the color distribution of GCs is expected
to be bimodal. This model predicts that the blue GC system shows
no rotation and a high velocity dispersion, while the red GC
system shows some rotation depending on the degree of dissipation
in the collapse. As the galactocentric distance increases up to
90\arcsec, the stellar velocity dispersion in M60 approaches to
the value of 200 km s$^{-1}$ \citep{debruyne01} that is comparable
to or smaller than the velocity dispersion of the blue GCs. This
might support this model, as \citet{forbes97} pointed out in the
case of NGC 1399 and M87. 
However, this model disagrees with the results that blue GC systems in several gEs including M60 
  in Table~\ref{tab-gesum} show larger than or comparable rotation
  amplitude to that of the red GC system.

\citet{cote98} proposed the dissipationless accretion model; the
red GCs formed in a dissipational collapse, while the blue GCs
were subsequently captured from other galaxies through mergers or
tidal stripping. This model predicts a bimodal color distribution
of GCs in gEs and a uniform color distribution in dwarf
ellipticals. However, there is an example of low-luminosity
elliptical galaxy (e.g., NGC 1427) that shows a bimodal color
distribution \citep{forte01}. Since the blue GCs are captured from
other galaxies, they are expected to show extended spatial
distribution. \citet{richtler04} argued that if the scenario of
\citet{cote98} is correct, then the blue GCs are expected to have
radially biased orbits rather than the isotropic or tangentially
biased orbits, and are also expected to show no rotation. Although
the kinematic data for the outer GCs are not complete to date,
current data for the blue GCs in Table~\ref{tab-gesum} show
negligible rotations and no signs of radially biased orbits, which
is not consistent with the scenario of \citet{cote98}.

Since the classical models do not give quantitative predictions
  concerning the kinematic properties of the GC system,
  it is instructive to compare the observational results
  with those in the recent numerical simulations.
Although there are some numerical simulations of the GC system focusing on
  the color distribution \citep{bea02},
  the spatial distribution \citep{moore06,bf06}, and
  the mass-metallicity relation for the blue GCs \citep{bekki07},
  there are few simulation results that
  can be compared directly with observational results summarized in Table \citep{bekki05,kra05}

In a pioneering work, \citet{bekki05} numerically investigated the kinematics
  of the GC system in E/S0 galaxies formed from a dissipationless merging of spiral galaxies.
They presented the kinematic properties such as rotation and velocity dispersion
  of the pre-existing MPCs and MRCs
  for several merger configurations (e.g., pair and multiple mergers).
For the rotation, both MPCs and MRCs show larger rotation amplitudes in the outer region ($R>2R_{\rm eff}$)
  than those in inner region ($R\sim R_{\rm eff}$), regardless of the merger configuration.
Interestingly, observational data in Figure \ref{fig-gerot} show that
  the rotation amplitudes for the blue GCs do not change as the galactocentric distance increases.
However, those for the red GCs increase marginally from the inner
region to the outer region
  if we neglect the point of NGC 4636 (filled pentagon), as expected in the simulation.
For the velocity dispersion, the MPCs show slightly larger the central velocity dispersion
 than the MRCs, indicating that the MPCs are dynamically hotter than the MRCs.
Moreover, the VDPs for both MPCs and MRCs
  are decreasing as the galactocentric distance increases in all major merger models,
  while those are sometimes very flat in multiple merger models.
The kinematic data in Figure \ref{fig-gesigresi} show that
  the VDPs might be different depending on the galaxy (e.g., flat for M49 and increasing for M87),
  and the difference of VDPs between blue and red GCs appears to exist.
Although the latter is not expected in the simulation, the former is consistent
  with the simulation, implying different merger histories depending on the galaxies.

\citet{kra05} studied the formation of the GCs in a Milky Way-size galaxy
  using a gas dynamics cosmological simulation with an adaptive refinement tree code.
However, they had to stop the simulation at $z\sim3.3$ due to the limited computational resources.
Later, \citet{gp06}, using a separate collisionless $N$-body simulation described in \citet{kra04} up to $z=0$,
  calculated the GC orbits for the similar galactic system.
They found that, at present, the GC orbits are isotropic in the
inner 50 kpc region from the galactic center,
  but radial in the outer region.
Although it is difficult to exam the observational data at large distance ($>50$ kpc)
  because of insufficient GC samples at that region,
  the isotropic orbit in the inner region
  is roughly consistent with the kinematic data of gEs (see Table \ref{tab-gesum}).

On the other hand, \citet{ves03} modeled a dynamical evolution of the M87 GC system 
  with different initial conditions for the mass function, 
  the spatial distribution, and the velocity anisotropy of the GC system, 
  following the evolution of a model GC system given by \citet{ves00,ves01}. 
In their result, for a two-slope power-law initial mass function of the GCs 
  corresponding to the mass function of the old clusters, 
  a flat radial profile for the mean mass of the GCs obtained from the observation 
  can be acquired from the simulation with any velocity anisotropy of the GC system. 
However, for a power-law initial mass function of the GCs 
  that is regarded as that of the young clusters, 
  the observed flat radial profile for the mean mass of the GCs 
  can be obtained from the simulation only with a strong radial anisotropy of the GC system. 
This initial radial anisotropy is much more radially biased 
  than the observed one in the GC systems of gEs 
  (having mostly isotropic velocity ellipsoids as seen in Table \ref{tab-gesum}), 
  raising a problem for a scenario that the young GC system in mergers like the Antennae 
  may evolve into the GC system seen in gEs.

By viewing the formation of the M60 GC system in the context of kinematic
  predictions of above models, the strong rotations of the blue and red GC
  system are consistent with the simulation of \citet{bekki05}.
However, the larger velocity dispersion of the red GCs compared to
the blue GCs in M60 is not
  consistent with the predictions of several formation models. M60 might be
  a particular case among gEs due to a possible interaction with the
  companion galaxy, NGC 4647.
While they were regarded as a non-interacting system (e.g., \citealt{sandage94}),
  recent observational results suggest evidences of current interaction between two:
  (1) the morphological structure of NGC 4647 is clearly asymmetric \citep{koopman01};
  (2) the stellar kinematic study shows that the inner region of M60
  has a strong rotational support compared to other gEs,
  and has an asymmetric rotation curve \citep{pinkney03,debruyne01};
  (3) an X-ray filament that extends to the north-eastern edge of M60 is seen \citep{randall06}.
  (4) young luminous star clusters or associations in NGC 4647 are found \citep{lee07}.
However, \citet{pierce06} found no obvious signs of a recent starburst,
  interaction or merger by estimating the ages and metallicities from the spectra of M60 GCs.
Thus, it appears that the interaction between two started very recently,
  and did not affect the old M60 GC system significantly.
The spatial coverage of observed GCs in \citet{pierce06} is not enough to study the
  interacting region between M60 and NGC 4647 (see Figure \ref{fig-vmap}).
Therefore, to understand the formation of M60 GC system in terms of an interaction
  between galaxies, it is important to obtain a large spectroscopic
  sample of GCs at larger galactocentric distance with high S/N ratio
  enough to determine the age and metallicity.

\section{Summary}

Using the photometric and spectroscopic database of 121 GCs (83 blue GCs and 38 red GCs) in the gE
M60 (NGC 4649) in the Virgo cluster, we have investigated the kinematics of the GC system of this galaxy.
Our primary results are summarized below:

\begin{enumerate}
\item Similar to the case of M87 GCs \citep{cote01}, the entire, blue
and red GC subsamples of M60 show significant overall rotations.
The rotation axes are nearly perpendicular to the line connecting
M60 and its companion NGC 4647.

\item Both of the velocity dispersion about the mean velocity and about the best fit rotation
curve of the red GCs are marginally larger than those of the blue GCs. This implies that the red GC
system might be dynamically hotter than the blue GC system unlike the GC systems in other gEs.

\item Comparison of observed stellar and GC velocity dispersion profiles with those
  calculated from the stellar mass profile
  showed that the mass-to-light ratio is not constant,
  but should be increased as the galactocentric distance increases,
  indicating the existence of an extended dark matter halo in M60.

\item  Using the X-ray mass profile, the number density distribution of GCs,
and the observed VDP of GCs, we have determined the velocity ellipsoids of the M60 GC system.
The entire GC system in M60 appears to have a tangentially biased velocity ellipsoid unlike the
GC systems in other gEs. Two subsamples have a different velocity anisotropy: the blue GCs show
a modest tangentially biased velocity ellipsoid, while the red GCs show
a modest radially biased or an isotropic velocity ellipsoid.

\item We have compared the kinematics of the M60 GC system in this study with the
results of other GC systems in gEs. As a whole, the
rotation-corrected velocity dispersion of the blue GCs is similar
to or larger than that of the red GCs unlike the M60 GC system.
The rotation of the GC system in gEs is not negligible in contrast
to the traditional view, though the details of rotation amplitudes
and rotation axes need to be investigated further. The entire
sample of each GC system appears to have an isotropic velocity
ellipsoid, while the subsamples do not show unified velocity
anisotropy.

\end{enumerate}

In conclusion, the GC systems in gEs have common kinematic
  properties such as velocity dispersion, while the rotation and the
  velocity anisotropy show diverse results. 
These kinematic properties are not fully explained by any current models.
We need extensive kinematic studies of other GC systems over
  larger galactocentric distances with various environmental effects
  to understand the diversity of the kinematics, and the formation
  and evolution of the GC system in gE. Moreover, it is desirable to
have more elaborate model predictions for the kinematics of the GC
system in gE including the velocity anisotropy.

\clearpage

\acknowledgments 
We would like to thank anonymous referee for useful comments.
The authors are grateful to the staff members of
the CFHT for their warm support during our observations. M.G.L. is
in part supported by the ABRL (R14-2002-058-01000-0).



\clearpage

\clearpage


\begin{deluxetable}{cccrrrrrr}
\tablewidth{0pc} 
\tablecaption{Kinematics of the M60
Globular Cluster System\label{tab-m60kin}}

\tablehead{
\colhead{$R$} & \colhead{$\langle R\rangle$} & \colhead{$N$} &
\colhead{$\overline{v_p}$} & \colhead{$\sigma_p$} &
\colhead{$\Theta_0$} & \colhead{$\Omega R$} &
\colhead{${\sigma}_{p,r}$} & \colhead{$\Omega R$/${\sigma}_{p,r}$}
\nl \colhead{(arcsec)} & \colhead{(arcsec)} & \colhead{} &
\colhead{(km s$^{-1}$)} & \colhead{(km s$^{-1}$)} &
\colhead{(deg)} & \colhead{(km s$^{-1}$)} & \colhead{(km
s$^{-1}$)} & \colhead{} }

\startdata
\multicolumn{9}{c}{Entire GCs: 121 Clusters with 1.0 $\le$ $(C-T_1)$
$<$ 2.4} \nl
~32--533 & 218 & 121 & $1073^{+22}_{-22}$ & $234^{+13}_{-14}$
 & $225^{+12}_{-14}$ & $141^{+50}_{-38}~$ & $217^{+14}_{-16}$
 & $0.65^{+0.27}_{-0.22}$  \nl
~32--200 & 116 & 60 & $1046^{+37}_{-35}$ & $252^{+15}_{-26}$
     & $212^{+18}_{-25}$ & $137^{+83}_{-48}~$ & $230^{+17}_{-29}$
     & $0.60^{+0.40}_{-0.28}$  \nl
200--533 & 319 & 61 & $1098^{+33}_{-29}$ & $217^{+15}_{-21}$
     & $241^{+16}_{-13}$ & $156^{+47}_{-38}~$ & $199^{+15}_{-23}$
     & $0.78^{+0.30}_{-0.28}$  \nl
\multicolumn{9}{c}{} \nl \multicolumn{9}{c}{Blue GCs: 83 Clusters
with 1.0 $\le$ $(C-T_1)$ $<$ 1.7} \nl
~32--533 & 228 & 83 & $1086^{+27}_{-25}$ & $223^{+13}_{-16}$
     & $218^{+16}_{-23}$ & $130^{+62}_{-51}~$ & $207^{+15}_{-19}$
     & $0.63^{+0.35}_{-0.30}$  \nl
~32--200 & 121 & 43 & $1076^{+37}_{-41}$ & $230^{+19}_{-28}$
     & $216^{+18}_{-33}$ & $139^{+119}_{-69}~$ & $199^{+21}_{-30}$
     & $0.70^{+0.67}_{-0.45}$  \nl
290--533 & 343 & 40 & $1097^{+39}_{-35}$ & $216^{+17}_{-22}$
     & $221^{+28}_{-26}$ & $121^{+51}_{-35}~$ & $212^{+19}_{-27}$
     & $0.57^{+0.29}_{-0.24}$  \nl
\multicolumn{9}{c}{} \nl \multicolumn{9}{c}{Red GCs: 38 Clusters
with 1.7 $\le$ $(C-T_1)$ $<$ 2.4} \nl
~32--533 & 197 & 38 & $1040^{+48}_{-42}$ & $258^{+21}_{-31}$
     & $237^{+18}_{-19}$ & $171^{+58}_{-46}~$ & $240^{+20}_{-34}~$
     & $0.71^{+0.30}_{-0.29}$  \nl
~32--200 & 102 & 17 & $961^{+72}_{-78}$ & $282^{+25}_{-45}$
     & $193^{+40}_{-37}$ & $139^{+94}_{-44}$ & $281^{+25}_{-45}~$
     & $0.49^{+0.38}_{-0.24}$  \nl
200--533 & 274 & 21 & $1100^{+55}_{-46}$ & $219^{+31}_{-55}$
     & $249^{+21}_{-18}$ & $218^{+83}_{-68}~$ & $186^{+21}_{-40}~$
     & $1.17^{+0.58}_{-0.62}$  \nl
\enddata

\end{deluxetable}
\clearpage


\begin{deluxetable}{ccccccccccccc}
\tablewidth{0pc} 
\tabletypesize{\scriptsize}
\tablecaption{Gas Temperature and Number Density\label{tab-xrayfit}}

\tablehead{

Data & \multicolumn{4}{c}{$T(r)$} & & \multicolumn{3}{c}{$n(r)$ ($i\le1$)} & & \multicolumn{3}{c}{$n(r)$ ($i\le2$)}  \\
\cline{2-5} \cline{7-9} \cline{11-13}
 & $T_m$      & $r_m$ & $r_{ot}$ & q & &$n_0(1)$   & $r_0(1)$ &  p(1) & & $n_0(1)$, $n_0(2)$ & $r_0(1)$, $r_0(2)$ & p(1), p(2)\\
 & ($10^7$ K) & (kpc) & (kpc)    &   & &(cm$^{-3}$)& (kpc)    &       & & (cm$^{-3}$) &(kpc)     &     }

\startdata

T97\tablenotemark{~a}     & 1.16 & 37.32 & 23.03 & $-$0.003 & & 0.02 & 2.60 & 1.45 & & ... & ... & ...  \\
T97+T86\tablenotemark{~b} & ... & ... & ... & ...       & & 0.09 & 0.70 & 1.41 & & 0.07, 0.0009 & 0.97, 16.83 & 1.69, 2.22  \\
R06\tablenotemark{~c}     & 7.47 & 19.49 & 17.34 & 0.19 & & 0.14 & 0.81 & 1.65 & & 0.06, 0.05 & 0.81,  1.61 & 3.08, 1.71  \\
BM97\tablenotemark{~d}    & 0.9~ & 24.28 & 24.28 & 0.0~ & & ... & ... & ... & & 0.1, 0.0014 & 0.9, 18.21 & 1.8, 3.0  \\

\enddata
\tablenotetext{a~}{\citet{trinchieri97}.}
\tablenotetext{b~}{\citet{trinchieri97} and \citet{trinchieri86}.}
\tablenotetext{c~}{\citet{randall06}.} \tablenotetext{d~}{Values
taken from \citet{brighenti97}.}
\end{deluxetable}
\clearpage


\begin{deluxetable}{ccrrcrr}
\tablewidth{0pc} 
\tablecaption{Giant Elliptical Galaxy Samples\label{tab-gesample}}

\tablehead{ \colhead{Galaxy} & \colhead{$M_V$\tablenotemark{~a}} &
\colhead{v$_{sys}$\tablenotemark{~b}} &
\colhead{$R_{\rm eff}$\tablenotemark{~c}} &
\colhead{$\epsilon$\tablenotemark{~d}} &
\colhead{PA$_{\rm minor}$\tablenotemark{~c}} &
\colhead{Distance\tablenotemark{~e}} \nl

\colhead{} & \colhead{} & \colhead{(km s$^{-1}$)} &
\colhead{(kpc)} & & \colhead{(deg)} & \colhead{(Mpc)}}

\startdata

M60 & $-$22.44 & 1056 & 9.23 & 0.21 & 15 & 17.3 \nl
M87 & $-$22.62 & 1307 & 7.66 & 0.125 & 69 & 17.2\nl
M49 & $-$22.83 & ~997 & 9.97 & 0.175 & 65 & 17.1 \nl
NGC 1399 & $-$21.95 & 1442 & 14.55 & 0.099 & 20 & 20.0 \nl
NGC 5128 & $-$21.66 & ~541 & 6.02 & 0.224 & 125 & 4.2  \nl

NGC 4636 &$-$21.43 & ~906 & 6.37 & 0.256 & 58 & 14.7 \nl
\enddata

\tablenotetext{a~}{Absolute $V$ magnitude based on $B_T$,
$(B-V)_T$ \citep{devaucouleurs91}, $A_V$ \citep{sch98}, and
distance adopted in this study.}
\tablenotetext{b~}{M60 (This study), M87, M49 \citep{smith00}, NGC 1399 \citep{richtler04}, NGC
5128 \citep{hui95}, and NGC 4636 \citep{schuberth06}.}
\tablenotetext{c~}{Position angle of the minor axis. M60 \citep{lee07}, NGC 5128 \citep{dufour79}, and other galaxies \citep{kim06}.}
\tablenotetext{d~}{M60 \citep{lee07}, NGC 5128 \citep{devaucouleurs91} and other galaxies \citep{kim06}.}
\tablenotetext{e~}{M60, M87, and M49 \citep{mei07}, and other galaxies \citep{tonry01}.}
\end{deluxetable}
\clearpage


\begin{deluxetable}{ccrrrrrcc}
\tablewidth{0pc} 
\tablecaption{Global Kinematic
Properties of GCs in gEs\label{tab-gekin}}

\tablehead{ \colhead{Galaxy} & \colhead{GC} & \colhead{$N$} &
\colhead{$\overline{v_p}$} & \colhead{$\sigma_p$} &
\colhead{$\Theta_0$} & \colhead{$\Omega R$} &
\colhead{${\sigma}_{p,r}$} & \colhead{$\Omega R$/${\sigma}_{p,r}$} \nl \colhead{} & \colhead{}
& \colhead{} & \colhead{(km s$^{-1}$)} & \colhead{(km s$^{-1}$)} &
\colhead{(deg)} & \colhead{(km s$^{-1}$)} & \colhead{(km s$^{-1}$)} & \colhead{} }

\startdata
M60 & EGC & 121& $1073^{+22}_{-22}$ & $234^{+13}_{-14}$ & $225^{+12}_{-14}$ & $141^{+50}_{-38}~$ & $217^{+14}_{-16}$
 & $0.65^{+0.27}_{-0.22}$  \nl
    & BGC &  83& $1086^{+27}_{-25}$ & $223^{+13}_{-16}$ & $218^{+16}_{-23}$ & $130^{+62}_{-51}~$ & $207^{+15}_{-19}$
     & $0.63^{+0.35}_{-0.30}$  \nl
    & RGC &  38& $1040^{+48}_{-42}$ & $258^{+21}_{-31}$ & $237^{+18}_{-19}$ & $171^{+58}_{-46}~$ & $240^{+20}_{-34}~$
     & $0.71^{+0.30}_{-0.29}$  \nl
\multicolumn{9}{c}{} \nl
M87 & EGC & 276& $1333^{+ 25}_{-23}$&$  414^{+ 15}_{-18}$&$   68^{+ 11}_{-12}$&$  172^{+ 39}_{-28}$&$  399^{+ 15}_{-18}$&$  0.43^{+0.12}_{-0.09}$ \nl
    & BGC & 158& $1341^{+ 36}_{-33}$&$  425^{+ 22}_{-25}$&$   59^{+ 17}_{-17}$&$  181^{+ 57}_{-44}$&$  414^{+ 22}_{-26}$&$  0.44^{+0.16}_{-0.13}$ \nl
    & RGC & 118& $1324^{+ 39}_{-37}$&$  400^{+ 25}_{-28}$&$   79^{+ 16}_{-17}$&$  165^{+ 53}_{-33}$&$  380^{+ 24}_{-27}$&$  0.43^{+0.17}_{-0.12}$ \nl
\multicolumn{9}{c}{} \nl
M49 & EGC & 263& $ 973^{+ 20}_{-18}$&$  322^{+ 14}_{-17}$&$  106^{+ 44}_{-45}$&$   54^{+ 50}_{-23}$&$  321^{+ 14}_{-17}$&$  0.17^{+0.17}_{-0.08}$ \nl
    & BGC & 159& $ 954^{+ 32}_{-27}$&$  352^{+ 21}_{-25}$&$  102^{+ 36}_{-37}$&$   92^{+ 71}_{-35}$&$  349^{+ 21}_{-24}$&$  0.27^{+0.22}_{-0.12}$ \nl
    & RGC & 104& $ 999^{+ 31}_{-25}$&$  276^{+ 19}_{-23}$&$  182^{+ 53}_{-50}$&$   11^{+ 79}_{-83}$&$  278^{+ 19}_{-23}$&$  0.04^{+0.29}_{-0.30}$ \nl
\multicolumn{9}{c}{} \nl
NGC 1399 & EGC & 435& $1442^{+ 15}_{-14}$&$  323^{+ 11}_{-13}$&$  307^{+ 50}_{-46}$&$   31^{+ 43}_{-48}$&$  326^{+ 11}_{-13}$&$  0.10^{+0.14}_{-0.15}$ \nl
         & BGC & 216& $1445^{+ 26}_{-22}$&$  359^{+ 17}_{-21}$&$  261^{+ 45}_{-52}$&$   69^{+ 68}_{-29}$&$  364^{+ 18}_{-21}$&$  0.19^{+0.20}_{-0.09}$ \nl
         & RGC & 219& $1439^{+ 19}_{-17}$&$  285^{+ 16}_{-19}$&$    0^{+ 40}_{-36}$&$   46^{+ 53}_{-39}$&$  288^{+ 16}_{-19}$&$  0.16^{+0.19}_{-0.15}$ \nl
\multicolumn{9}{c}{} \nl
NGC 5128 & EGC & 210& $ 536^{+  9}_{ -8}~$&$  129^{+  5}_{ -7}~$&$  184^{+ 23}_{-26}$&$   30^{+ 16}_{-14}$&$  129^{+  5}_{ -7}~$&$  0.23^{+0.14}_{-0.13}$ \nl
         & BGC & 127& $ 526^{+ 11}_{-11}$&$  126^{+  7}_{ -8}~$&$  168^{+ 27}_{-47}$&$   25^{+ 22}_{-35}$&$  129^{+  7}_{ -7}~$&$  0.19^{+0.18}_{-0.29}$ \nl
         & RGC &  83& $ 552^{+ 16}_{-15}$&$  133^{+  9}_{-11}$&$  191^{+ 40}_{-75}$&$   47^{+ 43}_{-54}$&$  132^{+  9}_{-11}$&$  0.36^{+0.35}_{-0.44}$ \nl
\multicolumn{9}{c}{} \nl
NGC 4636 & EGC & 172& $ 899^{+ 16}_{-14}$&$  207^{+ 10}_{-11}$&$  275^{+ 47}_{-50}$&$   29^{+ 35}_{-14}$&$  207^{+ 10}_{-11}$&$  0.14^{+0.18}_{-0.08}$ \nl
         & BGC &  96& $ 901^{+ 23}_{-20}$&$  207^{+ 14}_{-15}$&$  234^{+ 56}_{-52}$&$   19^{+ 49}_{-53}$&$  207^{+ 14}_{-15}$&$  0.09^{+0.25}_{-0.26}$ \nl
         & RGC &  76& $ 896^{+ 27}_{-24}$&$  208^{+ 12}_{-17}$&$  289^{+ 40}_{-42}$&$   50^{+ 46}_{-21}$&$  208^{+ 12}_{-18}$&$  0.24^{+0.23}_{-0.12}$ \nl

\enddata

\end{deluxetable}
\clearpage


\begin{deluxetable}{cccccc}
\tabletypesize{\footnotesize}
\tablewidth{0pc}
\tablecaption{Summary of Global Kinematic Properties of
              GCs in gEs\label{tab-gesum}}
\tablehead{
\colhead{Galaxy} &
\colhead{Velocity Dispersion} &
\colhead{GC} &
\colhead{Rotation} &
\colhead{Rotation Axis} &
\colhead{$\beta_{cl}$\tablenotemark{~a}}
} \startdata
M60 & $\sigma_{p,r}(BGC)\lesssim\sigma_{p,r}(RGC)$
      & EGC & Strong & Minor Axis & modest tangential \nl
    & & BGC & Strong & Minor Axis & modest tangential \nl
    & & RGC & Strong & None  & modest radial/isotropic \nl
\multicolumn{6}{c}{} \nl
 M87 & $\sigma_{p,r}(BGC)\gtrsim\sigma_{p,r}(RGC)$ & EGC & Strong & Minor
Axis & isotropic  \nl
    &  & BGC & Strong & Minor Axis & modest tangential  \nl
    &  & RGC & Strong & Minor Axis & modest radial  \nl
\multicolumn{6}{c}{} \nl M49 &
$\sigma_{p,r}(BGC)>\sigma_{p,r}(RGC)$  & EGC & Weak & None &
isotropic  \nl
    &  & BGC & Modest & None & closely isotropic \nl
    &  & RGC & Weak & Major Axis & closely isotropic  \nl
\multicolumn{6}{c}{} \nl NGC 1399 &
$\sigma_{p,r}(BGC)>\sigma_{p,r}(RGC)$ & EGC & Weak & Major Axis &
closely isotropic  \nl
    & & BGC & Weak & Major Axis & closely isotropic  \nl
    & & RGC & Weak & Minor Axis & closely isotropic  \nl
\multicolumn{6}{c}{} \nl NGC 5128 &
$\sigma_{p,r}(BGC)\simeq\sigma_{p,r}(RGC)$ & EGC & Modest & None & $-$  \nl
    & & BGC & Weak & None & $-$  \nl
    & & RGC & Modest & Major Axis & $-$  \nl
\multicolumn{6}{c}{} \nl NGC 4636 &
$\sigma_{p}(BGC)\simeq\sigma_{p}(RGC)$ & EGC & Weak & None & $-$  \nl
    & & BGC & Weak & Minor Axis & modest tangential  \nl
    & & RGC & Modest & None & $-$  \nl

\enddata
\tablenotetext{a~}{M60 (This study), M87 \citep{cote01}, M49 \citep{coteetal03},
NGC 1399 \citep{richtler04}, and NGC 4636 \citep{schuberth06}.}
\end{deluxetable}

\normalsize
\clearpage

\end{document}